%% file: main.tex
\documentclass[a4paper,11pt]{article}
\pdfoutput=1 % if your are submitting a pdflatex (i.e. if you have
\usepackage[shortlabels]{enumitem}
\usepackage{pst-all}	
\usepackage{physics}
\usepackage{blkarray}
\usepackage{mathtools}
\usepackage{booktabs}
\usepackage{jcappub} % for details on the use of the package, please
\usepackage[table,xcdraw]{xcolor}
\usepackage[T1]{fontenc} % if needed
\usepackage{cleveref}
\usepackage{tabularx} % for 'tabularx' env. and 'X' col. type
\usepackage[normalem]{ulem}
\usepackage{multirow}
\usepackage{tikz}
\usetikzlibrary{matrix, fit, positioning,arrows.meta,arrows, ,decorations.pathreplacing,shapes.geometric,backgrounds}

\title{Cosmological constraints from a joint DESI DR1 Full-Shape and DR2 BAO}

\include{authorlist}

\newcommand{\Mpch}{\,h^{-1}\,{\rm Mpc}}
\newcommand{\hMpc}{\,h\,{\rm Mpc}^{-1}}

\newcommand{\lcdm}{$\Lambda$CDM}
\newcommand{\wowacdm}{$w_0w_a$CDM}
\newcommand{\nulcdm}{$\nu\Lambda$CDM}
\newcommand{\nuwowacdm}{$\nu w_0w_a$CDM}
\newcommand{\olcdm}{$o\Lambda$CDM}
\newcommand{\onulcdm}{$o\nu\Lambda$CDM}
\newcommand{\owowacdm}{$o w_0w_a$CDM}

\newcommand{\lya}{Ly$\alpha$}

\abstract{
We present a cosmological analysis combining full-shape (FS) clustering measurements from the Dark Energy Spectroscopic Instrument (DESI) DR1 with baryon acoustic oscillation (BAO) measurements from DESI DR2. To achieve a robust combination that accounts for the correlation between the two data releases, we employ the ShapeFit compression method and estimate the joint covariance using EZmocks. This compressed approach inherently mitigates the prior volume effects that have previously dominated Bayesian constraints from DESI data with minimal external priors. Consequently, we obtain—for the first time within a Bayesian framework—reliable DESI-only constraints on extensions to $\Lambda$CDM using only a Big Bang Nucleosynthesis prior on the baryon density and a wide prior on the spectral index. In flat $\Lambda$CDM, we find $\Omega_m = 0.3035 \pm 0.0085$, $h = 0.6876 \pm 0.0059$, and $\sigma_8 = 0.822 \pm 0.034$. For the $w_0 w_a$CDM dynamical dark energy model, we measure $w_0 = -0.49 \pm 0.25$ and $w_a = -1.52 \pm 0.77$, improving constraints by $\sim 30\%$ relative to the analogous DR1 measurement and reducing the discrepancy with $\Lambda$CDM to $1.4\sigma$ when compared to BAO only analyses. We also report competitive limits on the sum of neutrino masses and spatial curvature. This work demonstrates that the ShapeFit compression provides a prior-robust and computationally efficient pathway to constrain beyond-$\Lambda$CDM physics with large-scale structure.
}

\crefname{figure}{figure}{figures}
\Crefname{figure}{Figure}{Figures}

\begin{document}
\maketitle
\flushbottom

\section{Introduction}
\label{sec:intro}

The Dark Energy Spectroscopic Instrument (DESI) \cite{levi2013desi, Levi2019,DESI2016a.Science} has surveyed the sky with unprecedented precision,   adding valuable information %from different points 
on the late history of the Universe from redshifts $z=0.1$ to $z\sim$4, providing enhanced constraints on cosmology.  DESI is located at the 4-meter Mayall Telescope in Kitt Peak (Arizona) and is a robotic, fiber-fed and highly multiplexed spectroscopic surveyor capable of observing around 5000 objects over a 3$^\circ$ field \cite{DESI2016b.Instr, Corrector.Miller.2023, FiberSystem.Poppett.2024} simultaneously. DESI observes spectra of four different tracers: Bright galaxies (BGS) \citep{BGSPrelim.RuizMacias.2020,BGS.TS.Hahn.2023}, Luminous Red Galaxies (LRG), \citep{LRG.TS.Zhou.2023}, Emission Line Galaxies (ELG) \citep{ELGPrelim.Raichoor.2020,ELG.TS.Raichoor.2023} and quasars (QSO) \citep{QSOPrelim.Yeche.2020,QSO.TS.Chaussidon.2023} the latter of which are also used to measure the Lyman-$\alpha$ (\lya{}) forest feature in the spectra. 
DESI is expected to observe more than 17,000 square degrees of sky and 63 million spectra over the extended 8-year observation period \cite{Spectro.Pipeline.Guy.2023}, and to place leading constraints on dark energy, neutrino masses and primordial non-Gaussianity. The First Data Release (DR1), including spectra for more than 18 million unique targets, is now public \cite{DESI2024.I.DR1}.

One of the main objectives of the experiment is to map the Baryon Acoustic Oscillations (BAO) feature in each tracer's clustering, which can be used to constrain the Universe expansion history and in particular, to provide joint constraints on the matter density parameter $\Omega_m$ and the combination of the sound horizon at drag epoch and the Hubble constant, $r_{\rm d}H_0$, in a $\Lambda$CDM model. Measurements from Data Releases (DR) 1 and 2 \cite{DESI2024.I.DR1,DESI.DR2.DR2} %have already hinted at the likelihood of an 
hinted at a possible evolving Dark Energy component \citep{DESI2024.VI.KP7A,DESI.DR2.BAO.cosmo,DESI.DR2.BAO.lya} when combining BAO constraints with Cosmic Microwave Background (CMB) data for the latter and CMB + Supernovae (SNe) for the former. 

While the BAO feature is a powerful and extremely robust cosmological probe of the background evolution, additional valuable information is enclosed in the Full-Shape (FS) of the galaxy/quasar power spectrum, which can be used to probe, e.g., growth of structure, neutrino masses and modified gravity. The first DESI analysis of FS measurements \citep{DESI2024.V.KP5,DESI2024.VII.KP7B,KP7s1-MG} combined with BAO measurements, confirmed the preference for evolving dark energy when combined with CMB and SN data.  When analyzing DESI-only data (FS + BAO) with minimal external data, Bayesian constraints on models beyond $\Lambda$CDM can be unreliable due to the influence of prior volume effects. This is especially true for models with an evolving dark energy equation of state.

 FS analyses come in two ``flavors'': full modeling and compressed. In the full modeling (FM) approach, adopted in \citep{DESI2024.V.KP5,DESI2024.VII.KP7B,KP7s1-MG}, the two-point statistics are modeled using a perturbation-theory description of large-scale structure that includes a bias expansion, redshift space distortions, and non-linear matter evolution. This allows the full broadband clustering to be modeled and fitted to the data simultaneously with the BAO feature.  
 The Full-Shape analysis considered here is of the compressed kind, particularly the ShapeFit approach \citep{Brieden21}, which uses the large-scale shape information in a manner that can later be reliably interpreted in the light of various cosmological models. The use of ShapeFit FS information (along with a BBN and an $n_s$ prior) also allows us to obtain $H_0$ constraints without relying on the knowledge of the sound horizon standard ruler length \cite{Brieden2023}. 
 
At the moment of writing, for DESI DR1, both BAO and FS analyses are available; but for DESI DR2 only BAO is. Hence, the aim of this paper is twofold. First, we provide the first consistent and optimal combination of DESI DR1 Full-Shape (FS) clustering and DR2 Baryon Acoustic Oscillation (BAO) measurements. By leveraging the enhanced constraining power of DR2 BAO, we update the FS cosmological constraints with the latest available data, serving as an interim analysis ahead of the official DR2 FS release. %\footnote{Pre-DR2 publication expected in mid 2026}.
Second, we demonstrate that the ShapeFit compression approach significantly reduces prior volume effects compared to Full-Modeling, without requiring additional priors beyond those of the baseline analysis. This allows us to derive robust cosmological constraints in beyond-\lcdm{} models with minimal reliance on external data, thereby underscoring DESI's standalone contribution to the field.
 %Second, we take advantage of the fact that the ShapeFit approach to FS analyses greatly mitigates the prior volume effects of FM without introducing additional (or external) priors to the standard ones of the baseline analysis to provide constraints with minimal external data additions, thus highlighting DESI's contribution to the state of the art.

Previous works looked into strategies to avoid prior volume effects in a Bayesian framework \citep{Paradiso2024,Zhang2025,Ivanov2025,Akitsu2024}. However, these approaches often rely on using the data to inform the prior choice or generally using more informative priors (such as HOD-informed priors). Alternatively, other works \citep{Morawetz-frequentist} forgo the Bayesian assumption and provide frequentist constraints. Alternatively, other works have developed analytical covariance approximations to the pre-post reconstruction two point clustering \cite{Maus2026}, with the caveat that the approximation ignores survey geometry effects, though the effect of this seems negligible in \lcdm{}. 

%We provide first-ever Bayesian constraints on beyond \lcdm{} models, adding a minimal amount of information to DESI data, that is, using a BBN prior on the physical baryon density, $\omega_b$, and a wide Gaussian prior on $n_s$. 

This work is organized as follows. \Cref{sec:data} introduces the data used, both DESI and external, then \cref{sec:methods} describes the methods used to combine DR1 and DR2 data while accounting for their correlation as well as reviewing the BAO and ShapeFit analyses. \Cref{sec:results} shows and discusses the results in terms of sample combinations and cosmology. Finally, we conclude in \cref{sec:conclusions}.

\section{Data and Mocks}
\label{sec:data}

\definecolor{lightgray}{HTML}{EFEFEF} % Define the exact shade

\begin{table}[h]
\centering
\scalebox{0.9}{\begin{tabular}{lllcccc}
\toprule
Tracer & Redshift Range & Data Release & $N_{\text{tracer}}$ & $z_{\text{eff}}$ & $V_{\text{eff}}$ (Gpc$^3$) & Use \\
\midrule

\multirow{2}{*}{BGS} & \multirow{2}{*}{0.1 – 0.4} & \cellcolor{lightgray}DESI DR1 & \cellcolor{lightgray}300,017 & \cellcolor{lightgray}0.295 & \cellcolor{lightgray}1.7 & \cellcolor{lightgray} --- \\
                      &                             & DESI DR2 & 1,188,526 & 0.295 & 3.8 & BAO\\ 
\multirow{2}{*}{LRG1} & \multirow{2}{*}{0.4 – 0.6} & \cellcolor{lightgray}DESI DR1 &\cellcolor{lightgray} 506,905 & \cellcolor{lightgray}0.510 &\cellcolor{lightgray} 2.6 & \cellcolor{lightgray} FS \\
                      &                             & DESI DR2 & 1,052,151 & 0.510 & 4.9 & BAO\\ 
\multirow{2}{*}{LRG2} & \multirow{2}{*}{0.6 – 0.8} & \cellcolor{lightgray}DESI DR1 &\cellcolor{lightgray} 771,875 & \cellcolor{lightgray}0.706 &\cellcolor{lightgray} 4.0 & \cellcolor{lightgray} FS \\
                      &                             & DESI DR2 & 1,613,562 & 0.706 & 7.6 & BAO\\ 
\multirow{2}{*}{LRG3} & \multirow{2}{*}{0.8 – 1.1} & \cellcolor{lightgray}DESI DR1 & \cellcolor{lightgray}859,824 & \cellcolor{lightgray}0.920 & \cellcolor{lightgray}5.0 & \cellcolor{lightgray}FS\\
                      &                             & DESI DR2 & 1,802,770 & 0.920 & 9.8 & ---\\
\multirow{2}{*}{LRG3+ELG1} & \multirow{2}{*}{0.8 – 1.1} & \cellcolor{lightgray}DESI DR1 &\cellcolor{lightgray} 1,876,164 & \cellcolor{lightgray}0.930 & \cellcolor{lightgray}6.5 & \cellcolor{lightgray} ---\\
                           &                             & DESI DR2 & 4,540,343 & 0.930 & 14.8 & BAO\\ 
\multirow{2}{*}{ELG2} & \multirow{2}{*}{1.1 – 1.6} & \cellcolor{lightgray}DESI DR1 & \cellcolor{lightgray}1,415,687 &\cellcolor{lightgray} 1.317 &\cellcolor{lightgray} 2.7  & \cellcolor{lightgray} FS\\
                      &                             & DESI DR2 & 3,797,271 & 1.317 & 8.3 & BAO\\ 
\multirow{2}{*}{QSO} & \multirow{2}{*}{0.8 – 2.1} & \cellcolor{lightgray}DESI DR1 &\cellcolor{lightgray} 856,652 &\cellcolor{lightgray} 1.491 &\cellcolor{lightgray} 1.5 & \cellcolor{lightgray}FS\\
                      &                             & DESI DR2 & 1,461,588 & 1.491 & 2.7 & BAO\\ 
\multirow{2}{*}{Ly$\alpha$} & \multirow{2}{*}{1.77 – 4.16} & \cellcolor{lightgray}DESI DR1 &\cellcolor{lightgray} 709,565 & \cellcolor{lightgray}2.330 &\cellcolor{lightgray} --- & \cellcolor{lightgray} FS (AP) \\
                            &                             & DESI DR2 & 1,289,874 & 2.330 & --- & BAO\\
\bottomrule
\end{tabular}}
\caption{Statistics for each of the DESI tracer types used
for the DESI DR1 FS and DR2 BAO measurements presented in this paper. The last column shows whether the sample is used in this analysis and in what form, either BAO or Full-Shape (FS). The FS information from \lya{} sample from DR1 is used to provide constraints on the Alcock-Paczynski (AP) effect rather than to constrain growth as is the case for the other tracers.}
\label{tab:desisamples}
\end{table}

\subsection{DESI dataset}
The Dark Energy Spectroscopic Instrument (DESI) started taking data for the main survey in May 2021 and will run for a total of eight years, surveying approximately 17000 square degrees of sky from redshift 0.1 to 4.2. DESI can simultaneously observe 5 thousand objects using optical fibers that are robotically positioned, making it significantly more efficient than previous spectroscopic surveys \cite{FiberSystem.Poppett.2024,FocalPlane.Silber.2023}. 

DESI observations belong in one of two programs. The Bright program is observed during bright nights and corresponds to the nearby galaxies, this is the BGS \cite{BGS.TS.Hahn.2023}. The Dark program observes the other tracers (LRG, ELG and QSO, \cite{LRG.TS.Zhou.2023,ELG.TS.Raichoor.2023,QSO.TS.Chaussidon.2023}). These different tracers allow DESI to cover a wide redshift range and obtain measurements at different points in the expansion history of the Universe. \Cref{tab:desisamples} shows the redshift ranges that these different samples cover as well as their estimated effective redshift and volume.

By June 2022, DESI had already gathered the DR1 \cite{DESI2024.I.DR1} sample, which comprises one year of data and has already been made public. The DR1 contains more than 6 million objects used for the cosmological analyses (more than twice the number of tracers in BOSS \cite{BOSS,eBOSS1}) and covers 7500 square degrees, corresponding to around 44\% of the final DESI area. \Cref{tab:desisamples} shows the number of tracers for each sample for DR1. These different redshift bins are considered disconnected and uncorrelated.

The DR2 comprises observations from the first three years of data from DESI and was finalized in April 2024. It represents more than a twofold increase in the number of objects relative to DR1, with more than 14 million tracers observed and a footprint increase of about 60\%.

\subsection{EZmocks}
DR1 and DR2 are highly correlated and their combination must take this into account.
The computation of the cross-correlation between DR1 and DR2 data requires a suite of mock survey catalogs.
This analysis employs  those generated with the EZmock method \cite{Chuang2015EZmock, Zhao2021}, which combines the computational speed of the Zel'dovich approximation \cite{Zeldovich1970} for dark matter evolution with a flexible bias model and a probability distribution function (PDF) mapping scheme. Ref.~\cite{Baumgarten_2018} have found that matching the 2-point statistics and the PDF is enough to accurately reproduce the bispectrum and the two-point covariance, thus the EZmocks aim at matching the two-point function while its PDF mapping ensures matching PDFs. Its efficiency and flexibility led to its adoption as the baseline for generating covariance matrices in eBOSS Collaboration cosmological analyses \cite{Zhao2021} and in DESI DR1 \cite{KP3s8-Zhao}. Below, we provide a brief overview of how we generate and process EZmock catalogues to emulate the observed DESI galaxy samples; a complete description of the mock suite is available in \cite{KP3s8-Zhao}.

Our DR1 EZmock suite comprises 1000 mocks for the North Galactic Cap and an equal number for the South Cap, for each tracer. For every tracer and redshift snapshot, we generated 2000 independent $(6\,h^{-1}\mathrm{Gpc})^3$ simulation boxes with $1728^3$ dark matter particles from Gaussian initial conditions. The dark matter density fields in these boxes are translated to galaxy density fields using a bias model. This model is calibrated snapshot-by-snapshot to match the two-point clustering of the DESI Abacus HOD mocks \cite{Yuan2024, antoine2023}, which themselves were tuned to data from the DESI survey validation (SV) \cite{DESI2023b.KP1.EDR}. The boxes are subsequently projected into sky coordinates. The large box size obviates the need for volume replication to cover an entire galactic cap, thereby preventing a misestimation of cosmic variance. Finally, snapshots are concatenated into single lightcones, onto which we apply the DESI DR1 survey footprint and Fast Fiber Assignment \cite{KP3s11-Sikandar} to enhance the realism of the final catalogues.

For the upcoming official DR2 FS analysis, a new suite of mock catalogs based on the bias model from \cite{Coloma2024} will be produced. In this work, as an interim solution, we repurpose 100 of the existing EZmock boxes created for DR1 \cite{KP3s8-Zhao}. We reproject these onto the DR2 sky footprint, concatenate the snapshots into redshift bins, and apply the full Fiber Assignment procedure \cite{KP3s15-Ross}. Using a larger ensemble of mocks is neither necessary nor practical for our purposes, since %a smaller 
this set is sufficient to  estimate reliably the cross-correlation between DR1 and DR2 at the level of compressed parameters (see \cref{sec:methods_combination}).

\subsection{External datasets and priors}\label{sec:external}

We closely follow the external data choices of the official DESI DR2 BAO \citep{DESI.DR2.BAO.cosmo} and DR1 Full-Shape \citep{DESI2024.VII.KP7B} analyses. In the absence of CMB data, we adopt priors on the spectral index $n_s$ and the physical baryon density $\omega_{\rm b}$. 

For $\omega_{\rm b}$, we use constraints from Big Bang Nucleosynthesis (BBN) by \cite{Schonenberg2024_BBN}, which are derived from the predicted abundances of light elements and account for uncertainties in nuclear interaction cross-sections. Following previous DESI analyses, we adopt the joint constraint on $\omega_{\rm b}$ and the effective number of relativistic species $N_{\rm eff}$ from \cite{Schonenberg2024_BBN}, fixing $N_{\rm eff}=3.044$. The BBN prior on $\omega_{\rm b}$ is then given by\footnote{Hereafter, $\mathcal{N}(\mu, \Sigma^2)$ denotes a normal distribution with mean $\mu$ and (co)variance $\Sigma^2$.}:
\begin{equation}
    \mathcal{L}(\omega_{\rm b}) = \mathcal{N}\left(\begin{pmatrix}0.02196 \\ 2.944\end{pmatrix}, 
        \begin{pmatrix}4.03\times 10^{-7} & 7.30\times 10^{-5} \\ - & 4.53\times10^{-2}\end{pmatrix}\right).
\end{equation}

For the spectral index $n_s$, we use a normal distribution with mean and standard deviation from Planck (2018) \citep{planck_collaboration_planck_2018}. For our baseline conservative scenario, we inflate the uncertainty by a factor of 10 (denoted $n_{s10}$):
\begin{equation}
    \mathcal{L}_{n_{s10}}(n_s) = \mathcal{N}(0.9649, 0.042^2).
\end{equation}

Due to the importance that Supernovae datasets have shown in previous DESI analyses, we also include them in this work, though focusing mostly on the PantheonPlus (PP) SNe sample \citep{PantheonPlus,Brout2022PP} for simplicity. The PP compilation consists of 1,550 spectroscopically-confirmed SNe Ia spanning redshifts from 0.001 to 2.26.

Finally, when employing CMB data, we adopt the temperature and polarization spectra (TT, EE respectively) and the cross spectrum (TE) from Planck (2018) PR3 \citep{planck_collaboration_planck_2018}. More precisely, for large scales ($\ell < 30$) we adopt the \texttt{SimAll Commander} and for small scales we use the \texttt{Plik} likelihood. Alternatively, in the case of neutrino and curved cosmologies, we use the \texttt{CamSpec} likelihoods from Planck PR4 \citep{PlanckPR41,PlanckPR42} as a baseline, following the choices of the DR2 BAO analysis. We supplement the primary CMB likelihood with the reconstructed CMB lensing power spectrum. This reconstruction uses the connected 4-point function of the CMB temperature and polarization maps. Specifically, we adopt the lensing data from the combination of the NPIPE PR4 \textit{Planck} reconstruction \cite{plancklens} and Data Release 6 of the Atacama Cosmology Telescope (ACT-DR6) \citep{act1,act2,act3}.
In what follows we refer to Planck TT, TE , EE and ACT lensing as ``CMB''.

\section{Methods}
\label{sec:methods}

To compute the cross correlation between DR1 and DR2 data we must perform the full DR2 BAO analysis in a set for DR2-like mock catalogs. 
As we use the (DR2-reprojected) EZmocks, we expect some differences between these DR2 mocks and the clustering signal of the DR2 data. However, because the mocks are not used to obtain the diagonal elements of the covariance matrix but only to describe the cross-correlation coefficients between DR1 and DR2, these potential differences in the clustering are not expected to impact our results in any significant way.
%, because the mocks are not used to obtain the diagonal elements of the covariance matrix but only to describe the cross-correlation coefficients between DR1 and DR2.
The large-scale production of mocks targeted to the DR2 dataset is expected to occur in the near future within the official DESI DR2 analysis workforce and is beyond the scope of this paper.

Below we briefly summarize the BAO and ShapeFit analyses performed on the mocks, but refer the reader to the relevant papers \citep{DESI2024.III.KP4,DESI.DR2.BAO.cosmo,DESI.DR2.BAO.lya,DESI2024.V.KP5,DESI2024.VII.KP7B} and references therein for an in-depth explanation.

\subsection{BAO}
The official BAO analysis within DESI is performed in configuration space by modeling the position of the BAO peak in the reconstructed galaxy two-point correlation function (2PCF). The BAO dilation parameters along and across the line-of-sight are obtained by fitting the BAO peak positions in the monopole and quadrupole of this 2PCF, while marginalizing over the broadband shape using the splines method described in \cite{KP4s2-Chen}. It should be noted that the dilation parameters thus only affect the peak position and are independent of the ``smooth'' or broadband component. The methodology is fully described in \cite{DESI.DR2.BAO.cosmo,DESI.DR2.BAO.lya}.

In this work, we adopt the same reconstruction convention and setup as in \cite{DESI2024.III.KP4,KP4s4-Paillas,KP4s3-Chen} and have tested that the pipeline applied to the DR2 EZmocks reproduces the official DR2 BAO results when applied to the data. 

We use the \texttt{desilike}\footnote{\url{https://github.com/cosmodesi/desilike}} library to sample the posterior of the BAO in terms of compressed parameters, i.e. relative dilation BAO peak position $q_\parallel$, $q_\perp$, where
\begin{align}
    q_\parallel(z) &= \frac{H^{\rm fid}(z)}{H(z)}\frac{r_d^{\rm fid}}{r_d},\\
    q_\perp(z) &= \frac{D_M(z)}{D_M^{\rm fid}(z)}\frac{r_d^{\rm fid}}{r_d}.
\end{align}
Here, $H(z)$ is the Hubble parameter, $D_M(z)$ the comoving angular diameter distance, $r_d$ the sound horizon at the drag epoch, and the superscript ${(\rm fid)}$ denotes quantities computed at a given fiducial cosmology. This fiducial cosmology is the one used to produce the catalog in terms of comoving distances and to define the 2PCF template from which the dilation parameters are obtained. The $q_\parallel$ is often expressed in terms of the radial Hubble distance, $D_H(z)\equiv c/H(z)$, such that $q_\parallel(z) = (D_H(z)/r_d)/(D_H^{\rm fid}(z)/r_d^{\rm fid})$.

These two parameters are often re-parametrized in terms of the Alcock-Paczynski parameter, $q_{\rm ap}$, and the isotropic volume dilation parameter, $q_{\rm iso}$, such that,
\begin{eqnarray}
    q_{\rm iso}(z)&\equiv& [q_{\parallel}(z)q_{\perp}^2(z)]^{1/3}, \\
    q_{\rm ap}(z)&\equiv& \frac{q_\parallel(z)}{q_\perp(z)}.
\end{eqnarray}

The BAO likelihood is then a multivariate Gaussian  given by
\begin{equation}
    \log\mathcal{L}_{\rm BAO}(\Theta_{\rm BAO}) = -\frac{1}{2}\qty[\xi_{\rm data} - \xi_{\rm model}(\Theta_{\rm BAO})]^T\mathbf{C}_{\rm BAO}^{-1}\qty[\xi_{\rm data} - \xi_{\rm model}(\Theta_{\rm BAO})],
\end{equation}
where $\Theta_{\rm BAO}$ stands for the vector of BAO compressed parameters, which can be either $\qty[q_{\rm iso}]$, $\qty[q_{\rm iso},q_{\rm ap}]$ or $\qty[q_{\parallel},q_{\perp}]$ depending on whether the tracer has enough signal-to-noise ratio to include 2D BAO measurements and how these are parameterized \citep{DESI.DR2.BAO.cosmo, DESI.DR2.BAO.lya}. $\xi_{\rm data}$ denotes the measured 2PCF multipoles (monopole and quadrupole) and $\xi_{\rm model}$ denotes the model 2PCF corresponding to the fiducial cosmology template modified by the BAO compressed parameters; $\mathbf{C}_{\rm BAO}$ denotes the covariance matrix. We use the same scale range used in the DR1/2 BAO analyses of $s\in\qty[50, 150]\Mpch$ in bins of $\Delta s = 4\Mpch$.
%
%Due to the simplicity of the model 
We are able to sample the posterior sufficiently densely using MCMC for all the EZmocks without the help of an emulator.

\subsection{ShapeFit}
Alternatively to the main DESI DR1 Full-Shape measurement \citep{DESI2024.V.KP5,DESI2024.VII.KP7B}, which performs the so-called Full-Modeling of the galaxy power spectrum to draw cosmological constraints, in this work, we adopt a model-agnostic alternative, ShapeFit \citep{Brieden21}. ShapeFit describes the power spectrum by compressing the $k$-band powers into a smaller data vector containing the re-parametrization of the scale dilation parameters, ($q_{\rm iso}^{\rm SF}$, $q_{\rm ap}^{\rm SF}$)\footnote{We will distinguish between $q^{\rm SF}$ and $q^{\rm BAO}$ for the parameters extracted from two different type of analyses: the index BAO stands for measuring only the BAO feature on the 2PCF of post-reconstructed catalogues, whereas SF stands for the full-shape (in terms of ShapeFit) of the power spectrum from pre-reconstructed catalogues. The two types of measurements are therefore covariant, but do not contain the same type of information.}, the logarithmic growth of structure parameter, $f$ (also expressed as $df \equiv f / f_{\rm fid}$ ) and two shape parameters, $m$ and $n$ (also expressed as, $dm \equiv m - m_{\rm fid}$ and $dn\equiv n-n_{\rm fid}$). On the one hand, the shape parameters $m$ and $n$ parametrise a change in the slope of the power spectrum at different scales, such that,
\begin{equation}
    P'_{\rm lin}(\vb{k}) = P_{\rm lin}(\vb{k})\exp{\frac{m}{a}\tanh(a\ln(\frac{k}{k_p})) + n\ln(\frac{k}{k_p})},
\end{equation}
where $k_p \approx 0.03\hMpc$ is the pivot wavenumber and $a\equiv0.6$ is the amplitude parameter. These are chosen to capture relevant information on the cosmology dependence of the transfer function (e.g. pivoting around $k_{\rm eq}$). On the other hand, the logarithmic growth of structure parameter $f$ controls the strength of RSD. Along 
with the linear bias $b_1$, it determines the relative amplitude of multipoles via $\beta = f/ b_1$ (see \cite{DESI2024.V.KP5} for more details).

The ShapeFit approach mirrors the standard BAO analysis pipeline. Both methods first extract a small number of compressed observables from the data by scaling and shifting a fiducial power spectrum template. This compression step aims to be cosmology-agnostic. The resulting parameters are then mapped onto predictions of a specific cosmological model in a second, independent step. This separation of data compression from cosmological inference bypasses the computational cost of generating a new full-theory power spectrum for each model evaluation during parameter sampling.

While a significant advantage of this approach was originally its computational efficiency, this is now secondary thanks to the speed of modern EFTofLSS emulators. Instead, in this work, we demonstrate that the compressed parameter space of ShapeFit has a regularizing effect, effectively mitigating the prior volume effects that have otherwise dominated DESI Full-Shape constraints in the absence of external data.

Previous work has already demonstrated the equivalence of these two approaches under various model choices \citep{KP5s4-Lai}, notably \lcdm, \wowacdm{} and neutrino cosmologies, which will be treated here.

The ShapeFit likelihood can be written as
\begin{equation}
    \log\mathcal{L}_{\rm SF}(\Theta_{\rm SF}) = -\frac{1}{2}\qty[P_{\rm data} - P_{\rm model}(\Theta_{\rm SF})]^T\mathbf{C}_{\rm SF}^{-1}\qty[P_{\rm data} - P_{\rm model}(\Theta_{\rm SF})],
\end{equation}
where $\Theta_{\rm SF} = \qty[q_{\rm iso}, q_{\rm ap}, dm, df]$ is the compressed ShapeFit data vector. Similarly to the BAO analysis, there are nuisance parameters that are marginalized over. Notice that the FS analysis is done in Fourier space, as in the main DESI results. $P_{\rm data}$ denotes the data power spectrum and includes the monopole and quadrupole within $k\in \qty[0.02, 0.2]\hMpc$ in bins of $\Delta k = 0.005 \hMpc$. We refer the reader to \citep{Brieden21,DESI2024.V.KP5} for explicit expressions of the model power spectrum $P_{\rm model}$. We perform our analysis fixing the amplitude of the power spectrum, and re-interpreting $f$ as $f\sigma_{s8}$, where $\sigma_{s8}$ is the amplitude of the matter fluctuation smoothed at the scale of $8\,{\rm Mpc}h^{-1}$, in cosmological units fixed to the fiducial cosmology.\footnote{This means $\sigma_{s8}$ accounts for the variation in smoothing scale ($8\Mpch$) due to the dilations ($q_{\rm iso}$, $q_{\rm ap}$). This quantity is therefore kept fixed at every step in the chain and equal to the value in the fiducial cosmology: $\sigma_{s8}\equiv\eval{\sigma_8}_{\Theta_{\rm fid}}$. The choice does not alter the fits, only their interpretation.} Similarly, we only vary one of the shape parameters, $m$, while keeping the other, $n$, fixed, and re-interpret them as $m+n$. For more details, see sec 3.1 of Ref.~\cite{novell-masot25}.

\subsection{Full-Shape Lyman-$\alpha$}
In the redshift range $z > 2$, the Lyman-$\alpha$ (Ly$\alpha$) forest --the absorption pattern imprinted on quasar spectra by neutral hydrogen in the intergalactic medium-- provides the most powerful tracer of large-scale structure within the DESI Survey. Traditionally, BAO measurements from the Ly$\alpha$ forest have delivered precise constraints on the cosmic expansion rate by isolating the peak position as the sole observable \cite{DESI.DR2.BAO.lya}. However, the clustering of the Ly$\alpha$ forest contains a wealth of additional cosmological information beyond the BAO peak. In particular, the broadband shape of the correlation function offers further opportunity to extract Alcock-Paczynski information \cite{Cuceu2021,Cuceu2023}.

Ref.~\cite{Lya-fs} performed a full-shape analysis of the Ly$\alpha$ forest 3D correlation functions. By modeling not only the BAO peak but also the broadband shape of the correlations, they achieved a significant improvement in constraining power. What we call \lya{} FS here is the use of Alcock-Paczynski information extracted not only from the BAO peak position but from the broadband itself. The key advancement was the precise measurement of the Alcock-Paczynski effect from the broadband, which, when combined with the BAO peak measurement, yields a 2.4 times tighter constraint on the distance ratio $D_H / D_M$ at an effective redshift of $z_{\rm Ly\alpha} = 2.33$ compared to a BAO-only analysis of the same data. They provide combined measurements from DR1 Full-shape data and DR2 BAO. We define the likelihood as the Gaussian approximation with covariance

\begin{equation}
    \mathbf{C}_{\rm Ly\alpha-FS} \equiv \begin{blockarray}{ccc}
 & D_H(z_{\rm eff}) / r_{d} & D_M(z_{\rm eff}) / r_{d} \\
\begin{block}{c(cc)}
D_H(z_{\rm eff}) / r_{d} & 0.077 & -0.016 \\
D_M(z_{\rm eff}) / r_{d} & - & 0.38 \\
\end{block}
\end{blockarray},
\end{equation}
and mean $\mu_{\rm Ly\alpha-FS} = \qty(D_H(z_{\rm eff}) / r_{d}, D_M(z_{\rm eff}) / r_{d}) = \qty(8.646, 38.90)$

\subsection{Combination of DESI DR2 BAO and DR1 Full-Shape}
\label{sec:methods_combination}
We run our EZmocks through the DESI BAO pipeline introduced above in order to obtain mock measurements  ${\bf M}^{\rm BAO} = [\bf{q}^{\rm BAO}_{\rm iso}, \bf{q}^{\rm BAO}_{\rm ap}]$, which is a matrix of size $(N_{\rm mocks}\times N_{\rm params}) = (100 \times 2)$ for BAO. Similarly, we perform ShapeFit measurements on the phase-matched 100 DR1 EZmocks with FA to obtain  ${\bf M}^{\rm SF} = [{\bf q}^{\rm SF}_{\rm iso}, {\bf q}^{\rm SF}_{\rm ap}, {\bf dm}, {\bf df}]$ (size $100 \times 4$).

The advantage of the ShapeFit approach is that there are only 8 DR2$\times$DR1 cross-covariance matrix coefficients, which can indeed be estimated reliably using a few mock realizations. If we were to use the alternative Full Modeling approach as in \citep{DESI2024.VII.KP7B}, then the cross-covariance elements would be $N_{\rm BAO}\times N_{P(k)}$, which would require at least an order of magnitude more realizations, rendering the analysis untractable at this stage. We follow the methodology introduced first in \cite{GilMarin2022_comb}, who tested different approaches to combine pre-reconstruction Full-Shape measurements and post-reconstruction BAO analyses, concluding that using compressed parameters yields results fully consistent with the full data-level combination.

Using the best-fit parameter values for the mocks, we can concatenate the compressed data vectors in order to obtain measurement matrices ${\bf M}^{\rm BAO + SF} = [{\bf q}^{\rm SF}_{\rm iso}, {\bf q}^{\rm SF}_{\rm ap}, {\bf dm}, {\bf df}, \bf{q}^{\rm BAO}_{\rm iso}, \bf{q}^{\rm BAO}_{\rm ap}]$ of size $100\times 6$. We can then use these measurements matrices to compute $6\times6$ covariance matrices ${\bf C}^{\rm X} = ({\bf M}^{\rm X})^{T}({\bf M}^{\rm X})$. In the case of $\mathbf{C}^{\rm BAO + SF}$, we divide it in three relevant parts as
\begin{equation}
     \mathbf{C}^{\rm BAO + SF} = \begin{pmatrix}
   \mathbf{C}^{\rm SF} & (\mathbf{C}^{\rm BAO \times SF})^{T}\\\
\mathbf{C}^{\rm BAO \times SF} & \mathbf{C}^{\rm BAO}
 \end{pmatrix},
 \label{eq:block_matrix}
\end{equation}
where the $\mathbf{C}^{\rm BAO \times SF}$ is the relevant sub-block that is the key product of this work.

Further, we compute the associated correlation matrices $(\mathbf{r}^X)_{ij}  = (\mathbf{C}^X)_{ij} / \sqrt{(\mathbf{C}^X)_{ii}(\mathbf{C}^X)_{jj}}$, as this is what is actually extracted directly from the mocks. We write this matrix as
%Further, we normalize the variance $C_{ii}$ in order to replace it with the data-obtained variance. The quantities actually obtained from mocks are the correlation matrix elements $(\mathbf{r}^X)_{ij}  = (\mathbf{C}^X)_{ij} / \sqrt{(\mathbf{C}^X)_{ii}(\mathbf{C}^X)_{jj}}$. We write this matrix as
\begin{equation}
     \mathbf{r}^{\rm BAO + SF} = \begin{pmatrix}
   \mathbf{r}^{\rm SF} & (\mathbf{r}^{\rm BAO \times SF})^{T}\\\
\mathbf{r}^{\rm BAO \times SF} & \mathbf{r}^{\rm BAO}
 \end{pmatrix}.
 \label{eq:block_matrix_correlation}
\end{equation}

The diagonal blocks can also be estimated by measuring the matrices ${\mathbf{r}}^{\rm BAO}_{\rm data}$, ${\mathbf{r}}^{\rm SF}_{\rm data}$ from the posterior samples of the official BAO and ShapeFit measurements from the DESI data.

We explore three different approaches to the covariance  for the combined analyses:
\begin{enumerate}[A.]
    \item Using the full correlation from the mocks:
    \begin{equation}
             \mathbf{r}^{\rm BAO + SF}_{\rm A} = \begin{pmatrix}
   \mathbf{r}^{\rm SF}_{\rm mock} & (\mathbf{r}^{\rm BAO \times SF}_{\rm mock})^{T}\\\
\mathbf{r}^{\rm BAO \times SF}_{\rm mock} & \mathbf{r}^{\rm BAO}_{\rm mock}
 \end{pmatrix},
    \end{equation}
    which is the correlation matrix associated with the mock compressed covariance
    \begin{equation}
       \mathbf{C}^{\rm BAO + SF}_{\rm A} \equiv (\mathbf{M}^{\rm BAO + SF})^T(\mathbf{M}^{\rm BAO + SF}).
    \end{equation}
     \item Ignoring the cross-correlations and using only data covariances. This is an approximation presented only for illustrative purposes, as the cross-correlation should not be neglected.
    \begin{equation}
             \mathbf{r}^{\rm BAO + SF}_{\rm B} = \begin{pmatrix}
   \mathbf{r}^{\rm SF}_{\rm data} & \mathbf{0}\\\
\mathbf{0} & \mathbf{r}^{\rm BAO}_{\rm data}
 \end{pmatrix}\equiv \mathbf{r}^{\rm BAO + SF}_{\rm data}.
    \end{equation}
    \item Using the diagonal blocks obtained from the data posteriors in the official analyses and estimating the off-diagonal blocks from the mocks.
    \begin{equation}
             \mathbf{r}^{\rm BAO + SF}_{\rm C} = \begin{pmatrix}
   \mathbf{r}^{\rm SF}_{\rm data} & (\mathbf{r}^{\rm BAO \times SF}_{\rm mock})^{T}\\\
\mathbf{r}^{\rm BAO \times SF}_{\rm mock} & \mathbf{r}^{\rm BAO}_{\rm data}
 \end{pmatrix}\equiv \mathbf{r}^{\rm BAO + SF}_{\rm B} + \begin{pmatrix}
     \mathbf{0} & \mathbf{1}\\ \mathbf{1} & \mathbf{0} 
 \end{pmatrix}\otimes \mathbf{r}^{\rm BAO + SF}_{\rm A},
 \label{eq:matrix_combination}
    \end{equation}
    where the $\otimes$ operator denotes element-wise product and $\mathbf{1}$ ($\mathbf{0}$) is a matrix of 1 (0) in every element.
\end{enumerate}

These correlation matrices are then converted back into covariances using the data variance as
\begin{equation}
    (\mathbf{\hat{C}}^{\rm BAO + SF}_{\rm A,B,C})_{ij} = \sqrt{(\mathbf{C}^{\rm BAO + SF}_{\rm data})_{ii}(\mathbf{C}^{\rm BAO + SF}_{\rm data})_{jj}}(\mathbf{r}^{\rm BAO + SF}_{\rm A,B,C})_{ij}
\end{equation}

Using each of these covariances, we combine the measurements into a single likelihood, where the ``LEQ'' subscript refers to the tracers LRG, ELG, QSO, 
\begin{align}
    \log\mathcal{L}_{\rm LEQ} =& \sum_{i \in~ \rm LRG,ELG,QSO}\log\mathcal{L}_{i}(z_{\rm eff}, \Theta_{\rm cosmo}) \\=& -\frac{1}{2}\sum_{i \in \rm LRG,ELG,QSO}\qty[\hat{\Theta}_{i,\rm comp}^{*,\,\rm data} - \Theta_{\rm comp}^{\rm model}]^T\qty(\mathbf{\hat{C}}^{\rm BAO + SF}_{\rm A,B,C})_i^{-1}\qty[\hat{\Theta}_{\rm comp}^{*,\,\rm data} - \Theta_{i,\rm comp}^{\rm model}],
\end{align}
where the compressed vector $\Theta_{\rm comp} = \qty[\Theta_{\rm SF},\Theta_{\rm BAO}]$ is the concatenation of BAO and ShapeFit compressed parameter vectors and the $\Theta^{\rm model}_{\rm comp}$ is evaluated at the point $\Theta_{\rm cosmo}$ in the cosmological parameter space. The superscript $(*,\rm data)$ denotes the maximum \textit{a posteriori} point in compressed parameter space extracted from the official DESI posterior (see \cref{appendix:sf-choice} for details). We have also rerun the analyses using the mean of the posterior instead and find no significant differences. We elaborate on this in \cref{appendix:sf-choice}.

$\mathcal{L}_{\rm LEQ}$ is the main product of this work. Given the unavailability of BGS DR2-like mock catalogs we are unable to compute the $\mathbf{r}^{\rm BAO\times SF}$ quantities in the covariance and default to using the DR2 BAO-only likelihood $\mathcal{L}_{\rm BGS}$. Similarly, we take the Lyman-$\alpha$ forest into account using the DR2 BAO + DR1 Full-Shape likelihood $\mathcal{L}_{\rm Ly\alpha-FS}$. We combine these into the DESI likelihood
\begin{equation}
    \log\mathcal{L}_{\rm DESI} \equiv \log\mathcal{L}_{\rm LEQ} + \log\mathcal{L}_{\rm BGS} + \log\mathcal{L}_{\rm Ly\alpha-FS}.
\end{equation}

For cosmological inference, we employ \texttt{Cobaya} \cite{Cobaya} with \texttt{CAMB} \cite{CAMB} and use the external likelihoods implemented in the package for CMB and SNe.

Note that, among these A, B, and C options presented here, option C is the best at balancing the actual errors of the data, with the approximate cross-terms obtained from the mocks. Options A and B are only presented here as qualitative comparisons on how the results are affected by this choice (see later \cref{sec:performance_matrix}). On the one hand, B completely ignores the BAO and ShapeFit correlation in the scaling parameter, which we know by construction should exist, as SF partially use BAO information from the pre-reconstructed catalog. On the other hand, A assumes a fixed cross-term independent of the realization of the data. In fact, we know these terms do suffer from fluctuations, and they should be properly estimated for each of these realizations. Option C only fixes the off-diagonal cross-terms to be realization-independent; although this is strictly an approximation, we expect it to have little impact on the data. In order to assess such an impact, we later compare in \cref{fig:lcdm_cov_comp} the performance of applying each of these three methodologies to the DESI data.

\section{Results and Discussion}
\label{sec:results}

The main results of this work include the correlation (and covariance) for the compressed parameters between DESI BAO DR2 and DESI FS DR1 (\cref{sec:combination_results}) and the cosmological interpretation of their optimal combination within $\Lambda$CDM (\cref{sec:lcdm_constraints}) and several of its popular extensions (\cref{sec:w0wacdm_constraints,sec:nucdm_constraints,sec:olcdm_constraints,sec:late_constraints}).

\subsection{Combined correlation matrix}
\label{sec:combination_results}
\begin{figure}
    \centering
    \includegraphics[width=0.99\linewidth]{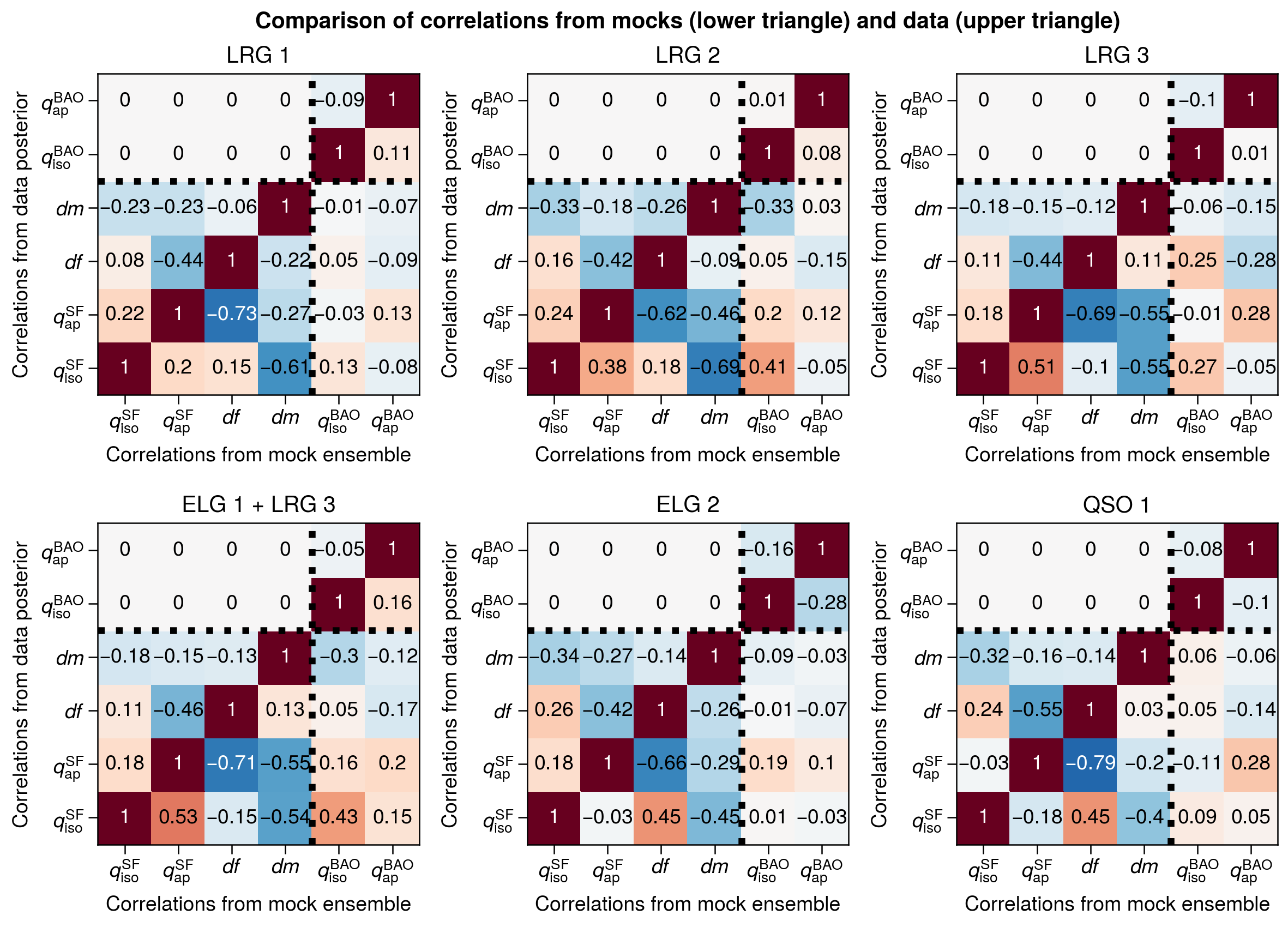}
    \caption{Correlation matrices of the compressed Full-Shape $\times$ BAO data vector. Top triangle: Matrix elements estimated from DR1(2) ShapeFit (BAO) posterior samples, i.e. without the cross-correlation terms ($\mathbf{r}^{\rm BAO+ SF}_{\rm B}$). Bottom triangle: Matrix elements estimated in this work using mock ensemble, thus including the cross correlations ($\mathbf{r}^{\rm BAO+ SF}_{\rm A}$).}
    \label{fig:correlation_matrices}
\end{figure}

\Cref{fig:correlation_matrices} shows the correlation matrices obtained from the data posteriors ignoring the cross correlation (upper triangle, $\mathrm{r}^{\rm BAO+SF}_{\rm B}$) \citep{DESI2024.V.KP5,DESI.DR2.DR2} and from the mocks (lower triangle, $\mathrm{r}^{\rm BAO+SF}_{\rm A}$). A first visual inspection shows that these matrices are similarly structured. The seemingly large differences between mock and posterior estimated cross terms (such as in the case of $q_{\rm ap}^{\rm SF}\times q_{\rm iso}^{\rm SF}$ for LRG3) are consistent with the scatter observed among the mocks. A relatively large scatter is expected from these parameter covariance elements, as it has been shown to depend on the particular realization \cite{GilMarin2022_comb}.  We do not explicitly show $\mathrm{r}^{\rm BAO+SF}_{\rm C}$ as it is built as a combination of the former two as detailed in \cref{eq:matrix_combination}\footnote{Available from \url{https://github.com/dforero0896/DESI1.5}}.

\begin{figure}
    \centering
    \includegraphics[width=0.99\linewidth]{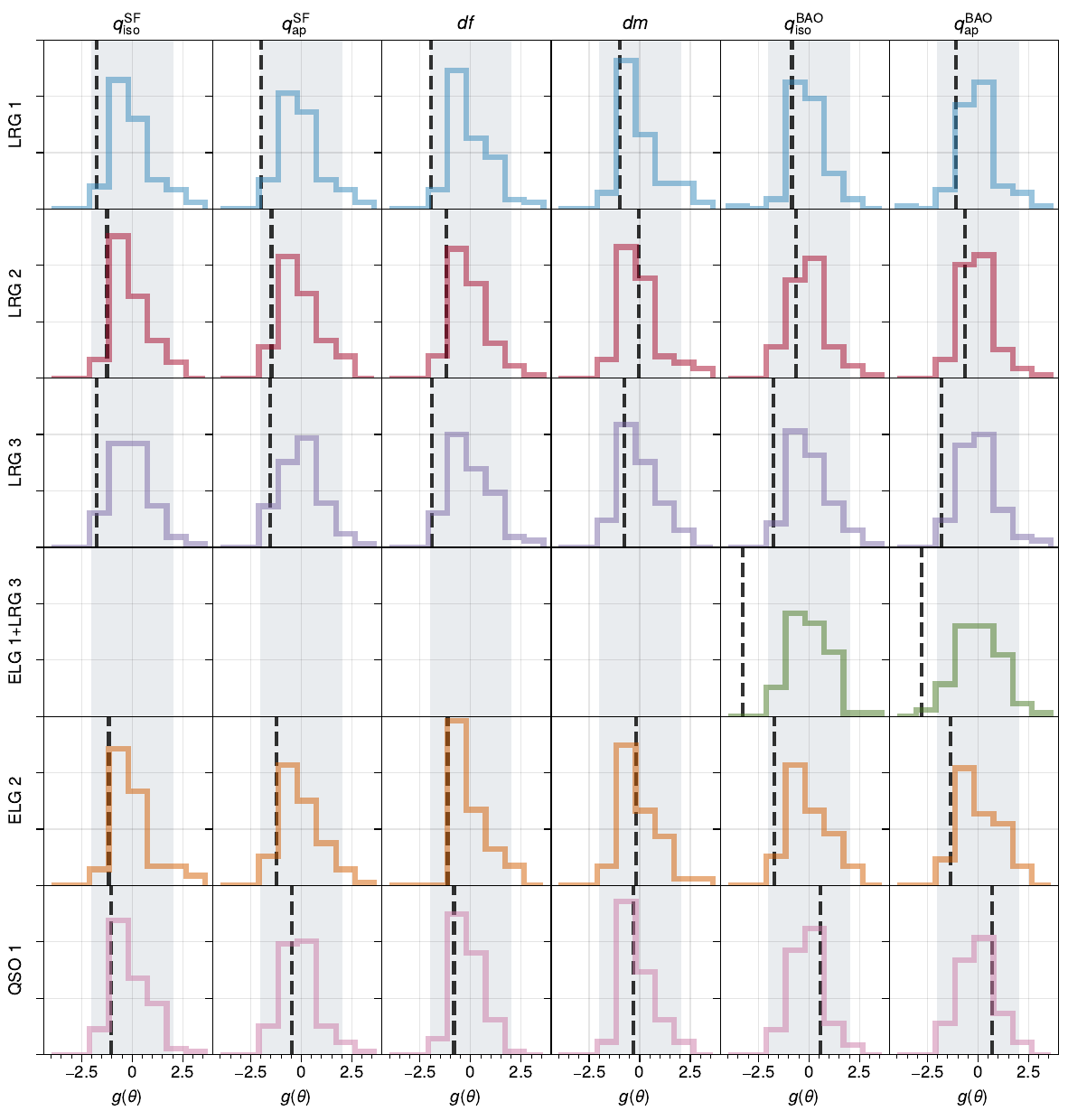}
    \caption{Distribution of errors for each of the six compressed parameters measured from the 100 EZmocks (histograms) compared to the actual data errors (vertical lines), for different tracers. The quantity shown corresponds to the $Z$-score component, $g(\theta) = \qty(\sigma_\theta - \bar{\sigma}_\theta) / \mathrm{std}(\sigma_\theta)$, where $\bar{\sigma}_\theta$ denotes the mean error on parameter $\theta$ over the EZmocks and $\mathrm{std}(\sigma_\theta)$  the corresponding standard deviation. Each panel corresponds to one parameter-sample combination, the shaded area shows the $2\sigma$ region. Given that the combined ELG1+LRG3 sample was not used in the FS analyses, these panels are empty.} \label{fig:validate_errors}
\end{figure}

One way to validate quantitatively that the mock-based covariance matrix estimates are a good description of the statistics of the data is to assess whether the data are a plausible draw from the underlying distribution of mocks.
We  thus compute the standardized measurement error (pull, $Z$-statistic) for each compressed parameter $\theta$ over the 100 EZmock realizations, defined as $g(\theta) = (\sigma_\theta - \bar{\sigma}_\theta) / \mathrm{std}(\sigma_\theta)$, where $\mathrm{std}(\sigma_\theta)$ is the standard deviation of $\sigma_\theta$. The distribution of these pulls for each sample and parameter is shown in \cref{fig:validate_errors}. While we expect some discrepancies between data and mocks in the original clustering statistics due to approximations in the EZmocks, we find that in the compressed parameter space ($\Theta_{\mathrm{comp}}$), the differences are not significant for most cases. The vertical lines in the figure panels show the corresponding values from the actual data catalog, and the gray band indicates the $2\sigma$ region around the expectation from the mocks.

We find that most data errors lie within $2\sigma$ of the expectation from the mocks and are located in bins containing some mock realizations, indicating these deviations are expected from the distribution of mocks. The strongest deviations originate from the combined ELG+LRG sample, which we attribute to two factors. First, there is a known mismatch between the covariance matrices derived from the EZmocks and those from the RascalC \cite{rascalC,rascal-jackknife,rascalC-legendre-3,2023MNRAS.524.3894R} method when the latter is tuned directly to the data \cite{KP4s6-Forero-Sanchez,KP3s15-Ross}, though this has been accounted for in the FS fits through the variance rescaling described in \cite{KP3s15-Ross}. Second, the EZmock galaxy populations have not been tuned to reproduce the observed cross-correlation between different tracers, leading to an imperfect match for this specific sample combination. Nevertheless, cosmological analyses performed with and without this combined sample yield no significant differences in the final parameter constraints. We therefore proceed to use the combined sample for all results presented in this work.

\subsection{Performance of  DR1$\times$DR2 matrices in \lcdm}\label{sec:performance_matrix}

\begin{figure}
    \centering
    \includegraphics[width=0.99\linewidth]{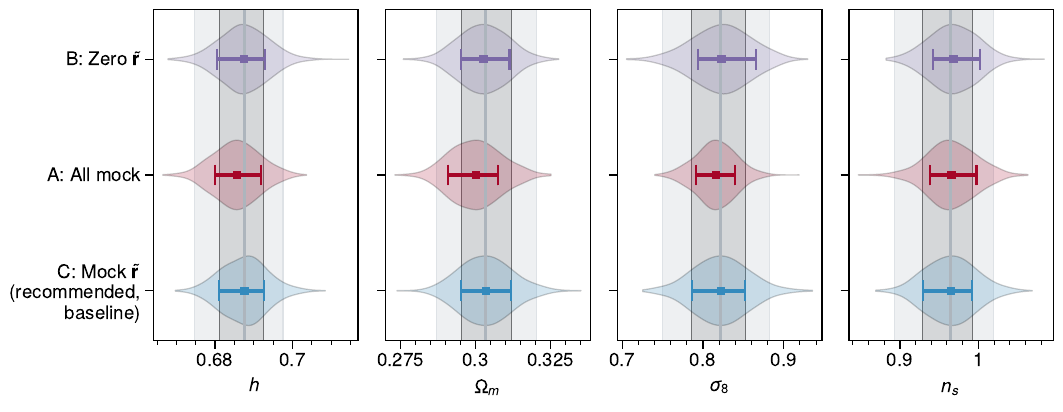}
    \caption{Comparison of different marginalised cosmological posteriors for different covariance choices (see section~\ref{sec:methods_combination}) within the \lcdm{} model. We show our baseline covariance matrix choice (C) at the bottom row and show its corresponding 1 and 2$\sigma$ limits as shaded grey areas and the mean as a vertical line, to facilitate the comparison with the other two approaches.}
    \label{fig:lcdm_cov_comp}
\end{figure}

\Cref{fig:lcdm_cov_comp} is a whisker plot for the posterior distributions of the relevant \lcdm{} cosmological parameters employing the covariance matrices constructed using the A, B, and C prescriptions described previously. 
Results for the parameter estimates are generally consistent with one another, which provides an upper bound for the potential systematic effects arising from fixing the off-diagonal cross-terms of option C to be realization-independent. Note that the errors estimated from the full EZmock-based parameter covariance (A) are smaller than those of the other two approaches. Ref. \cite{GilMarin2022_comb} shows that parameter cross covariances (even within the same analysis, i.e $q_{\rm iso}^{\rm SF}\times q_{\rm ap}^{\rm SF}$) can fluctuate significantly across realizations and lead to over- or under-estimates of the final parameter errors. The conservative choice is then to use such covariance elements from the data realization, i.e. the posterior-derived ones, as done in options A and C, contrary to option B.

In terms of the inclusion of DR1$\times$DR2 cross correlations, the terms $\mathbf{r^{\rm BAO\times SF}}$ have little effect on the cosmological inference results, at least when included in terms of compressed parameters, causing only minor shifts in $h$ and $\Omega_m$. This is likely due to the significantly more constraining power of DR2 BAO compared to DR1 FS, we do not expect this to be the case when the datasets have a similar constraining power.

Our previously-stated baseline choice, C, produces the most conservative (i.e., largest) errors in this specific case, although this behavior could potentially be different for other realizations. Moreover, this test shows that the differences in errors between C and B are not very large; therefore, the potential effect of fixing the off-diagonal terms to be realization-independent does not strongly impact the final estimated errors.

\subsection{Dataset nomenclature}\label{sec:names}
 In what follows, we refer to the main results of this paper, i.e. DR1 Full-Shape plus DR2 BAO analysis as ``DESI1.5''. This dataset is built up using the baseline covariance choice C, computed from the 100 EZmocks, as described above.
 
 In the next sections, we will often compare DESI1.5 with the official DESI analyses.  We refer to the Full-Shape plus BAO constraints from DR1 as ``DESI1''. DESI1 results can be presented using the Full-Modelling (FM) setup \cite{DESI2024.VII.KP7B}, hereafter DESI1 (FM), or the ShapeFit (SF) compression setup \cite{DESI2024.V.KP5}, hereafter DESI1 (SF). Unless both FM and SF variants are presented --in which case it is explicitly labeled-- DESI1 refers to the FM-based results from \cite{DESI2024.VII.KP7B}. Note that DESI1.5 is inherently only SF; for simplicity, we omit the (SF) specification. Finally, we refer to DR2 BAO only as ``BAO2'' \citep{DESI.DR2.BAO.cosmo}.

\subsection{\lcdm{} constraints}
\label{sec:lcdm_constraints}

We start by analyzing the new DESI1.5 results within the flat \lcdm{} model. We adopt the BBN prior on $w_b$, and the $n_{s10}$ prior, as described in section~\ref{sec:external}. 

\begin{figure}
    \centering
    \includegraphics[width=0.99\linewidth]{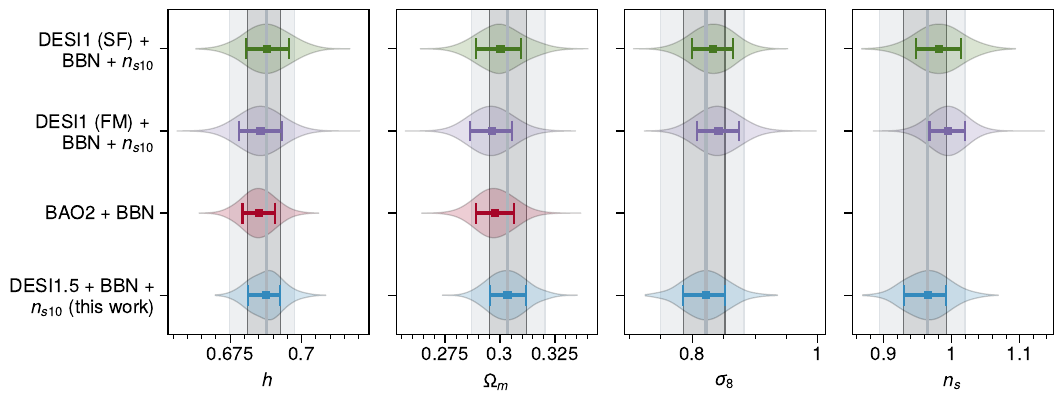}
    \caption{Comparison of our results (DESI1.5) to previous DESI1 and BAO2 analyses in the context of \lcdm. Constraints from DESI1 using FS constraints, both from FM and SF, are also shown for reference.}
    \label{fig:lcdm_sanitycheck}
\end{figure}

The results in terms of the 1D marginalized posteriors are displayed in blue in \Cref{fig:lcdm_sanitycheck}. We also display the BAO2 (red) and the DESI1, both Full-Modelling  (purple) and ShapeFit  (green)  constraints. Results for DESI1.5 are in very good agreement with both DESI1 and BAO2 analyses, while the error bars are comparable, the recovered central values for $h$ and  $\Omega_m$ are shifted by 0.4 and 0.7$\sigma$ compared to BAO2. In addition, we recover smaller values of $n_s$ (0.9$\sigma$) and $\sigma_8$ (0.58$\sigma$) than DESI1 with FM. This can be attributed to the 20\% broader $n_s$ constraint inherent to the ShapeFit approach, which allows for a larger $\Omega_m$ (smaller $n_s$). This behavior has already been found in previous work \citep{KP5s2-Maus, Maus23} where it has also been shown to disappear once tighter priors on $n_s$ are used or equivalently, when combined with CMB data. These shifts and the differences in $n_s$ are explored in detail in a forthcoming paper \cite{Asensio-SF}.

\begin{table}
    \centering
    \scalebox{0.7}{\begin{tabular}{p{3cm}cccc}
        \toprule
        Dataset & $\Omega_m$ & $h$ & $\sigma_8$ & $n_s$\\
\midrule 
\rowcolor[HTML]{EFEFEF}
BAO2 + BBN & $0.2978 \pm 0.0087$  & $0.6850 \pm 0.0058$  & -- & --\\ 
DESI1 (SF) + 
 BBN + $n_{s10}$ & $0.300 \pm 0.010$  & $0.6880 \pm 0.0077$  & $0.833 \pm 0.033$  & $0.982 \pm 0.034$ \\ 
\rowcolor[HTML]{EFEFEF}
DESI1 (FM) + 
 BBN + $n_{s10}$ & $0.2963 \pm 0.0095$  & $0.6856 \pm 0.0075$  & $0.842 \pm 0.034$  & $0.994 \pm 0.026$ \\ 
\textbf{DESI1.5 + BBN + $n_{s10}$} & \boldmath$0.3035 \pm 0.0085$  & \boldmath$0.6876 \pm 0.0059$  & \boldmath$0.822 \pm 0.034$  & \boldmath$0.965 \pm 0.032$ \\ 
%\rowcolor[HTML]{EFEFEF}
%\textbf{DESI1.5 + BBN + $n_{s10}$ - $r_d$} & \boldmath$0.302 \pm 0.011$  & \boldmath$0.693 \pm 0.026$  & \boldmath$0.823 \pm 0.035$  & \boldmath$0.960 \pm 0.041$ \\ 
        \bottomrule
                        \end{tabular}}
        \caption{\textbf{Cosmological constraints in \lcdm}. We add constraints from DESI1 analyses using FM and SF to illustrate the $n_s$ discrepancy and compare with our latest DESI1.5 analyses (highlighted in bold). We further include, for reference, DR2 BAO constraints. These results correspond to those shown also in \cref{fig:lcdm_sanitycheck}.}
        \label{tab:cosmo_params_lcdm}
    \end{table}

The results for the \lcdm{} parameters displayed in \Cref{fig:lcdm_sanitycheck} for the new DESI1.5 compilation (blue contours), are thus,
  \begin{equation}
    \begin{rcases}
    \Omega_m &= 0.3035 \pm 0.0085 \\
\sigma_8 &= 0.822 \pm 0.034 \\
h &= 0.6876 \pm 0.0059 \\
\end{rcases}
    \quad
    \text{DESI1.5 + BBN + $n_{s10}$} 
    \end{equation}
whose signal is dominated by BAO2 data in $\Omega_m$ and $h$ and DESI1 data in $\sigma_8$.
For completeness, we also display in \cref{tab:cosmo_params_lcdm} the results for DESI1 and BAO2, corresponding to those shown in \cref{fig:lcdm_sanitycheck}.

\subsection{\wowacdm{} Dynamical Dark Energy}
\label{sec:w0wacdm_constraints}
\begin{table}
    \centering
    \resizebox{\columnwidth}{!}{\begin{tabular}{p{3cm}cccccc}
        \toprule
        Dataset & $\Omega_m$ & $h$ & $w_0$ & $w_a$ & $\sigma_8$ & $n_s$\\
\midrule 
\rowcolor[HTML]{EFEFEF}
BAO2 + BBN & $0.354_{-0.017}^{+0.041}$  & $0.646 \pm 0.021$  & $-0.46 \pm 0.26$  & $-1.73_{-1.27}^{+0.33}$  & -- & --\\ 
DESI1 (SF) + BBN + $n_{s10}$ & $0.330 \pm 0.033$  & $0.674 \pm 0.033$  & $-0.67 \pm 0.33$  & $-1.3 \pm 1.0$  & $0.803 \pm 0.042$  & $0.951 \pm 0.038$ \\ 
\rowcolor[HTML]{EFEFEF}
\textbf{DESI1.5 + BBN + $n_{s10}$} & \boldmath$0.351 \pm 0.025$  & \boldmath$0.645 \pm 0.022$  & \boldmath$-0.49 \pm 0.25$  & \boldmath$-1.52 \pm 0.77$  & \boldmath$0.782 \pm 0.038$  & \boldmath$0.956 \pm 0.039$ \\ 
BAO2 + CMB & $0.352 \pm 0.021$  & $0.638 \pm 0.019$  & $-0.42 \pm 0.21$  & $-1.73 \pm 0.60$  & $0.783 \pm 0.016$  & $0.9661 \pm 0.0037$ \\ 
\rowcolor[HTML]{EFEFEF}
DESI1 (SF) + CMB & $0.326 \pm 0.027$  & $0.663 \pm 0.028$  & $-0.63 \pm 0.26$  & $-1.29 \pm 0.69$  & $0.806 \pm 0.024$  & $0.9658 \pm 0.0037$ \\ 
%DESI1 (FM) + CMB \cite{Morawetz-frequentist} & $0.331_{-0.032}^{+0.035}$  & $0.659 _{-0.031}^{+0.034}$    & $-0.58 _{-0.32}^{+0.33}$  & $-1.52_{-0.93}^{+0.88}$  & $0.812_{-0.27}^{+0.29} $ & ---\\ 
%\rowcolor[HTML]{EFEFEF}
\textbf{DESI1.5 + CMB} & \boldmath$0.356 \pm 0.020$  & \boldmath$0.634 \pm 0.017$  & \boldmath$-0.38 \pm 0.20$  & \boldmath$-1.82_{-0.53}^{+0.66}$  & \boldmath$0.778 \pm 0.015$  & \boldmath$0.9667 \pm 0.0035$ \\ 
        \bottomrule
                        \end{tabular}}
\caption{Cosmological parameters from \wowacdm{} dark energy parametrization from DESI1.5 combinations. As a reference, we cite the constraints from DR2 BAO and DESI1 (SF), showing negligible projection effects too. Our latest constraints (DESI1.5), highlighted in bold and also seen in the text, are the first constraints from DESI alone (including full-shape) and DESI + CMB with negligible projection effects within the Bayesian paradigm, thanks to our ShapeFit compression.}
\label{tab:w0wa_params}
                    \end{table}

\begin{figure}
    \centering
    \includegraphics[width=0.99\linewidth]{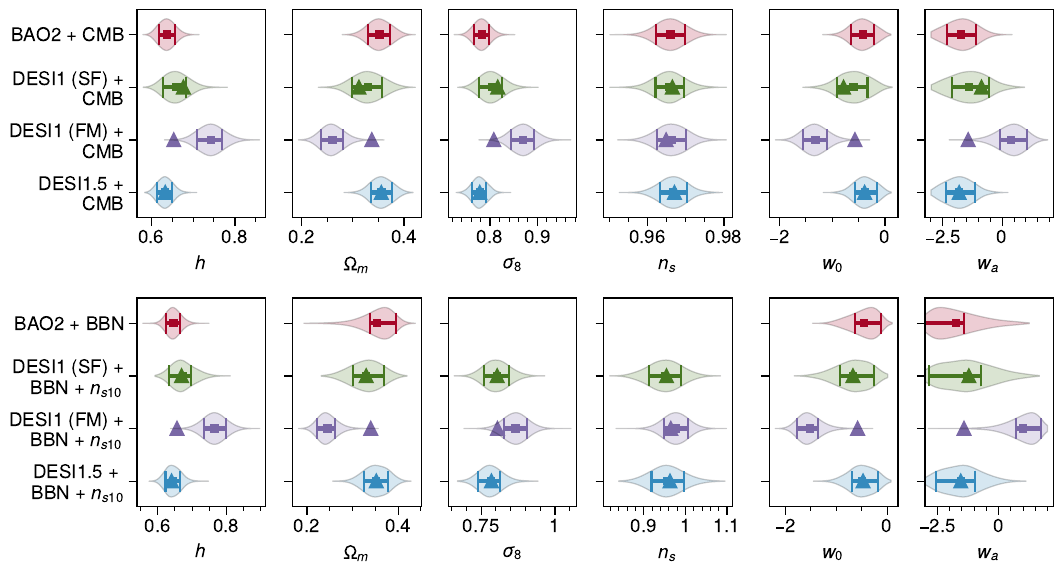}
    \caption{Marginalized 1D posteriors under \wowacdm{}. We also show the corresponding MAP values for each (FS) likelihood as a triangular symbol. The systematic shift between the MAP and the posterior in DESI1 (FM) shows the presence of significant prior volume effects in combinations with BBN and CMB. Our results, i.e. DESI1 (SF) and DESI1.5, do not show these effects as the MAP overlap the posterior mass. We plot the corresponding BAO constraints for reference.}
    \label{fig:projection_effects}
\end{figure}
                    
\begin{figure}
    \centering
    \includegraphics[width=0.49\linewidth]{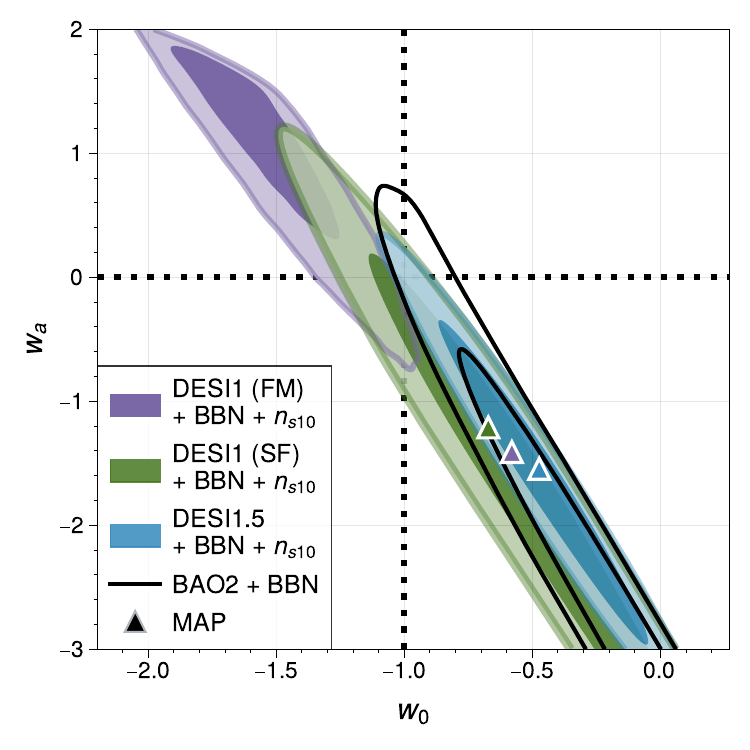}
    \includegraphics[width=0.49\linewidth]{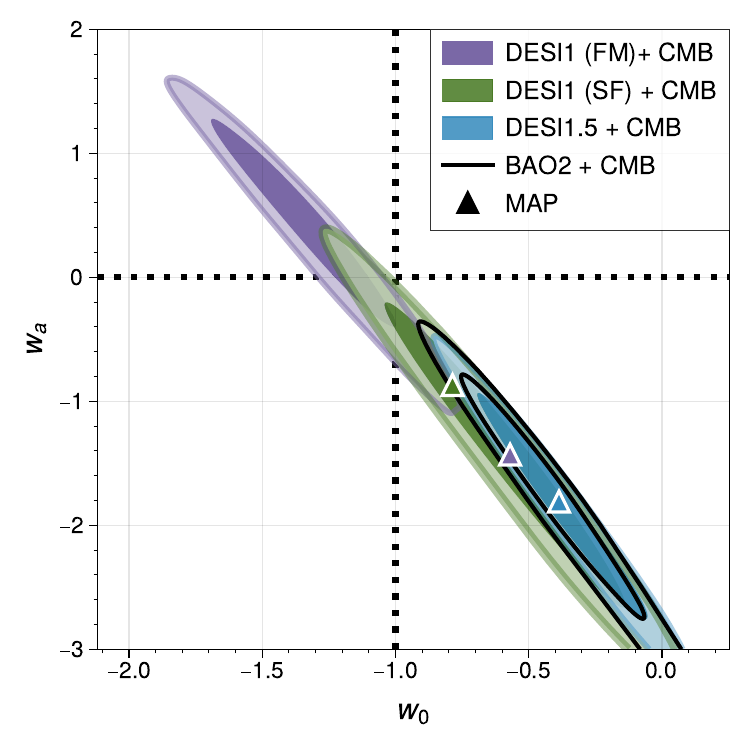}
    \caption{Constraints on dynamical dark energy. Left: DESI-only constraints using BBN and a loose prior on $n_s$. Purple and green contours show the posteriors from DESI1 Full-Modeling (FM) and ShapeFit (SF) analyses, respectively. The blue contour shows the combined DR1 FS + DR2 BAO (DESI1.5) result from this work. Triangles mark the maximum \textit{a posteriori} (MAP) estimates from the DESI1 (FM) analysis (taken from \cite{Morawetz-frequentist}) and this work's DESI1 (SF) and DESI1.5. Right: The same DESI constraints combined with CMB data. In both panels, the DESI1 (FM) contours exhibit strong prior volume effects, visible as a shift between the MAP point (triangle) and the contour. This effect is mitigated in DESI1 (SF) and DESI1.5, which remain consistent with the MAP estimates. The empty contours show the BAO results from DESI DR2 \cite{DESI.DR2.BAO.cosmo} for reference.}
    \label{fig:w0wacdm+cmb}
\end{figure}
%CMB row from frequentist
%\rowcolor[HTML]{EFEFEF}
%DESI1 (FM) + CMB \cite{Morawetz-frequentist} & $0.331_{-0.032}^{+0.035}$  & $0.659 _{-0.031}^{+0.034}$    & $-0.58 _{-0.32}^{+0.33}$  & $-1.52_{-0.93}^{+0.88}$  & $0.812_{-0.27}^{+0.29} $ & ---\\
We now turn to the evolving dark energy model through the popular parametrization in terms of the $w_0$ and $w_a$  equation of state parameters \cite{ChevallierPolarski,Linder03}. This is known to be a particularly interesting case for potential departures of $\Lambda$CDM model, as when the BAO2 dataset is combined with CMB and SNe data suggests that dark energy might not be a cosmological constant, $\Lambda$, but time-evolving; this preference for a dynamical dark energy model
over $\Lambda$ varied from 2.8 up to 4.2$\sigma$ depending on the  SNe sample selected\footnote{Some reanalysis of SNe samples have been performed since then \cite{DESreanalysis}, reducing the initial 4.2$\sigma$ preference for DESY5 published in \cite{DESI.DR2.BAO.cosmo}} \cite{DESI.DR2.BAO.cosmo}.

Working with the ShapeFit compressed parameters instead of the FM approach helps to dramatically reduce projection effects that arise during the posterior marginalization process \cite{DESI2024.VII.KP7B,Morawetz-frequentist}.

%allows us to circumvent the projection effects that biased the Bayesian contours in the earlier FM analysis. 

Projection effects occur when a complex, high-dimensional posterior distribution is marginalized over some parameters, leading to distortions that include shifts between the maximum of the marginal posterior and the maximum of the profiled posterior (the one obtained by maximizing a subset of parameters). When these misalignments are strong, can cause to misleading interpretations within the Bayesian framework \cite{Hadzhiyska2023,Tsedrik2025}. In the cosmological context, full-shape analyses require marginalization over a suite of (nuisance) parameters that are poorly constrained by the data, and occupy a lot of volume in parameter space where the size of the likelihood is not particularly large. Thus, the priors on those nuisance parameters are informative and dominate the marginalization process. In particular, even when selecting uniform priors, the exact parametrisation chosen to define the nuisance parameters (i.e. $b_1$ or $b_1\sigma_8$) affects the final marginalized posterior, and any priors (even when they are uniform) turn out to be informative.
We choose to work with the ShapeFit compressed variables (instead of the FM parameters), which have been shown to largely mitigate these projection effects and result in a set of cosmology results very insensitive to the priors on those nuisance parameters \cite{Brieden21}, thus removing most of the projection effects.

This can be appreciated in \cref{fig:projection_effects}, where we compare the marginalized posteriors for the BBN+$n_{s10}$ (lower rows) and CMB combinations (upper rows) obtained from DESI1 (FM), with those obtained from methods presented in this work, DESI1 (SF) and DESI1.5. In most cases, the DESI1 (FM) contours exhibit clear prior volume effects, visible as a systematic offset between the posterior distribution and the corresponding maximum a posteriori (MAP) estimate (triangle symbol). In contrast, our SF-based (for both DESI1 and DESI1.5) analyses yield MAP values that lie within the high-probability regions of their respective posteriors, demonstrating that projection effects are substantially mitigated by the ShapeFit compression.

A detailed view of the dark energy parameter constraints is shown in \cref{fig:w0wacdm+cmb}. On the $w_0$--$w_a$ plane, the DESI1 (FM) posterior (purple) is concentrated in the quadrant $w_0<-1, w_a>0$, while its MAP (matching triangle symbol) lies in the opposite quadrant ($w_0>-1, w_a<0$), well outside the $1\sigma$ credible region. The SF-based results (green and blue) produce posteriors that are consistent with their MAP estimates, confirming the robustness of the compression approach without requiring more restrictive priors. The empty black contours show the corresponding constraints without FS information for reference. It is clear on the left panel that adding FS makes constraints tighter, implying a significant information gain, even when the effective volume of the FS data is much smaller than BAO's. On the contrary, when adding CMB (right panel), DESI1.5 contours barely shrink relative to BAO2+CMB. This can be attributed to a minimal information gain from FS over CMB, the latter of which already constrains the matter transfer function far better than the former.

%This can be appreciated in \cref{fig:w0wacdm+cmb}, where the left panel shows the DESI1+BBN+$n_{s10}$ constraints from using FM and SF as well as the corresponding maximum \textit{a posteriori} (MAP) estimates from maximizing the DESI1 (FM) posteriors as reported in \cite{Morawetz-frequentist} and our own computations for DESI1 (SF) and DESI1.5. The FM contour lies in the second quadrant ($w_0<-1, w_a>0$) of the $w_0-w_a$ plane, whereas all the MAP lie in the fourth ($w_0>-1, w_a<0$). Conversely, the SF-based posterior is centered around these locations and fully consistent with the likelihood contours obtained by a frequentist analysis \cite{Morawetz-frequentist}, thus showing that prior volume effects are vastly mitigated. Similar conclusions can be drawn from the right panel of the same figure, where CMB information is included,  but is not sufficient to remove the projection effects in DESI1 (FM) posteriors. 

%\HGM{I am a bit confused about two things: 1) why I can only see a MAP for the purple. Should I assume that the green/blue MAPs are very close? If so, you should say these are not displayed for clarity, but they are very close to the purple ones. 2) I am not sure I see a qualitatively different behavior between blue in the right and in the left, to say that `it's not sufficient to remove the projection effects inherent in DESI1 FM"}
%\DF{I have now computed and added the MAP for each contour, except DESI1 (FM) which I just take from their paper. }

The absence of significant projection effects allows us to quote the Bayesian DESI1.5 + BBN + $n_{s10}$ constraints for the first time

\begin{equation}
    \begin{rcases}
    \Omega_m &= 0.351 \pm 0.025 \\
h &= 0.645 \pm 0.022 \\
\sigma_8 &= 0.782 \pm 0.038 \\
w_0 &= -0.49 \pm 0.25 \\
w_a &= -1.52 \pm 0.77 \\
\end{rcases}
    \quad
    \text{DESI1.5 + BBN + $n_{s10}$} 
    \end{equation}

In conclusion, adding DR2 BAO to DR1 FS, improves the constraints on $h$, $\Omega_m$ and DE parameters by $\sim30\%$ relative to the analogous DR1 analysis (i.e. DESI1 (SF)). DESI1.5 reduces the  ``tension''  with \lcdm{} to  $1.4\sigma$ compared to the $1.7\sigma$ observed from BAO2 alone \cite{DESI.DR2.BAO.cosmo}. These results are summarized in \cref{tab:w0wa_params}.

In combination with CMB, DESI1.5 yields:
 \begin{equation}
    \begin{rcases}
    \Omega_m &= 0.356 \pm 0.020 \\
h &= 0.634 \pm 0.017 \\
\sigma_8 &= 0.778 \pm 0.015 \\
w_0 &= -0.38 \pm 0.20 \\
w_a &= -1.82_{-0.53}^{+0.66} \\
\end{rcases}
    \quad
    \text{DESI1.5 + CMB} 
    \end{equation}

Compared to the BAO2+CMB constraints our DESI1.5+CMB analysis improves constraints by about $5\%$ while shifting means by less than 0.3$\sigma$.

Other approaches to mitigate projection effects have been presented in \cite{Zhang2025,Tsedrik2025}. Using the method presented in \cite{Tsedrik2025}, ref.~\cite{Chudaykin2025b} obtain FS+BAO+CMB constraints that are compatible with \lcdm{} at the $2\sigma$ level. While our constraints are comparable in size to theirs (even before including the bispectrum monopole), their recovered central values for  $w_0$, $w_a$ are 1.2 and 0.2$\sigma$ away from ours, respectively. These differences are not unexpected given differences in the analysis choices and the approximations made in \cite{Chudaykin2025}, in particular of neglecting the DR1-DR2 covariance and assuming Gaussian covariance matrices.

\subsection{Neutrino mass constraints}
\label{sec:nucdm_constraints}
%We perform analyses with neutrino cosmologies in order to constrain 
The sum of neutrino masses can be constrained from the neutrino mass effect on the linear growth rate of structure, the neutrino-suppression of the small-scale power spectrum and from the effect on the expansion history. In this work, we impose a physical $\sum m_\nu$ positive prior for our posterior sampling and consider both \lcdm{}, \wowacdm{} and spatially curved (\olcdm{}) backgrounds. To estimate where the $\sum m_\nu$ posteriors peak, we fit a Gaussian close to the prior edge and quote the fitted mean $\sum m_\nu^*$ as an approximation to the peak position.

BAO measurements alone, which exclusively trace the expansion history at relatively low redshifts, cannot directly constrain the neutrino masses. Instead, the BAO data helps break parameter degeneracies in combination with CMB observations, leading to tight constraints on $\sum m_\nu$. The addition of FS information provides sensitivity to the neutrino-induced suppression of small-scale power, enabling an upper limit to be placed on DESI data without CMB. Nevertheless, the strongest limits remain those from the joint analysis with CMB, where the complementary degeneracy directions yield the highest constraining power. For completeness and as a demonstration of the self-contained potential of large-scale structure, we also report DESI-only constraints in this work.

\subsubsection{\nulcdm{} constraints}
\begin{figure}
    \centering
    \includegraphics[width=0.49\linewidth]{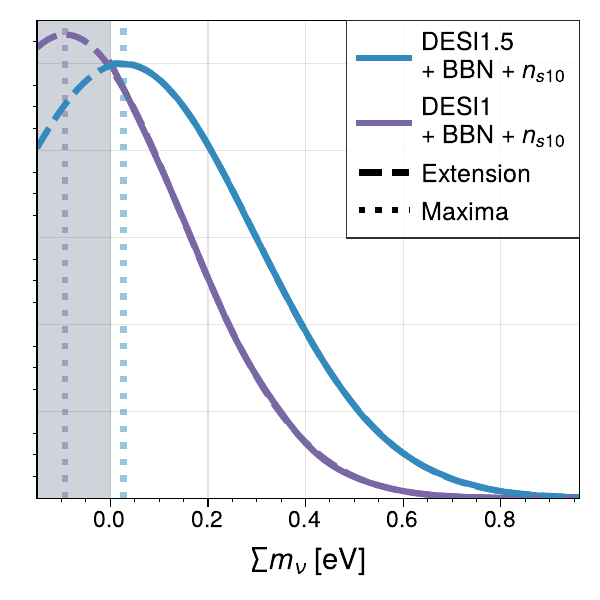}
    \includegraphics[width=0.49\linewidth]{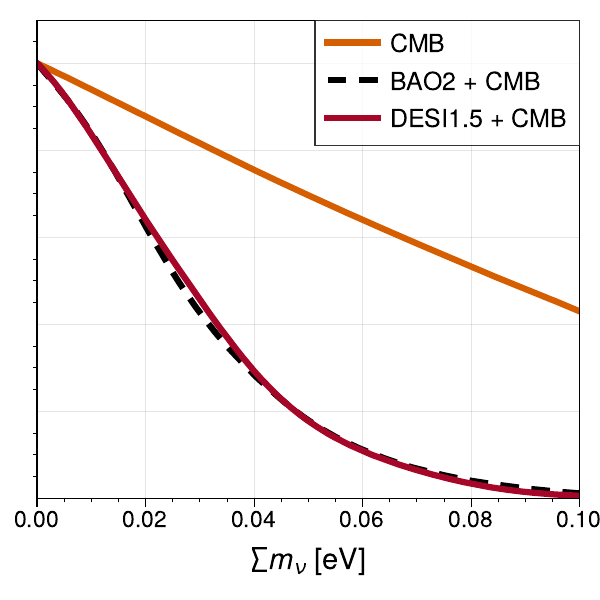}
    \includegraphics[width=0.75
    \linewidth]{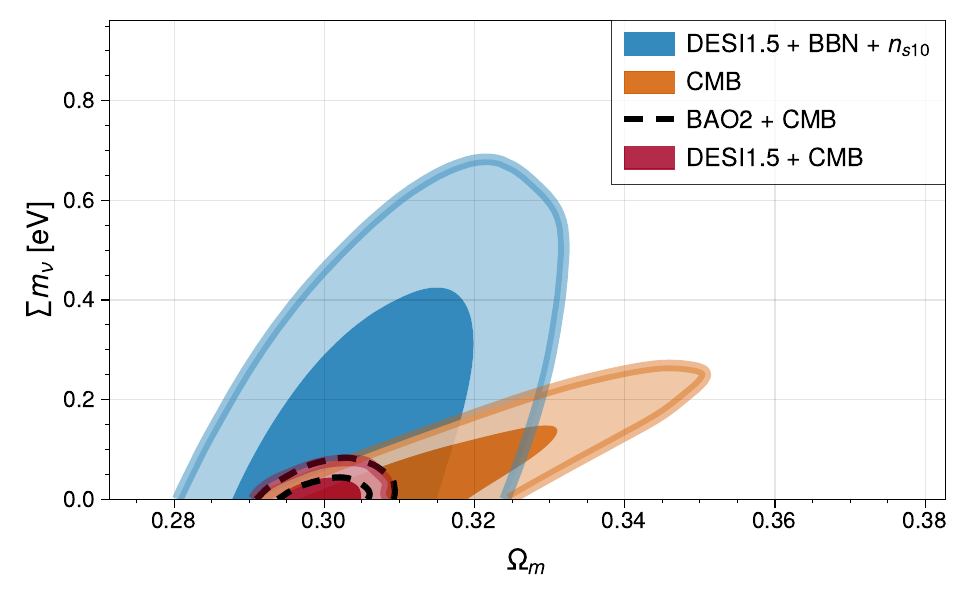}
    \caption{Neutrino mass constraints within \lcdm. Top left: DESI-only constraints from this work and DR1-only results. We extend the posteriors using Gaussian fits to show the location of their peak (dotted lines). Top right: DESI+CMB\texttt{[Camspec]} constraints from FS and DR2 BAO showing their consistency. CMB-only posterior added for reference. Bottom panel: 2-dimensional marginal posterior in the $\Omega_m$, $\sum m_\nu$ space showing the DESI1.5 constraints, and the degeneracy breaking with CMB that drives cosmological neutrino mass constraints.}
    \label{fig:nucdm}
\end{figure}

%We first perform 
Our baseline FS analyses, including the BBN and $n_{s10}$ priors with the full DESI data yields a 95\% confidence threshold on the neutrino mass of  
\begin{equation}
    \sum m_\nu < 0.5419 ~\mathrm{eV} \quad (95\%\text{ CI}), \quad \text{DESI1.5 + BBN + $n_{s10}$},
\end{equation}
an upper limit which is notably larger than the DR1 value quoted in \cite{DESI2024.VII.KP7B}. In order to understand this result,  the left panel of \cref{fig:nucdm} displays the marginalized posteriors for the sum of neutrino masses for both DESI1 (FM) and DESI1.5, both with the external priors on BBN and $n_{s10}$, as indicated.
%FS compared to the DR1-only analogous results. 
A Gaussian fit to the posterior (dashed line) has been added as a proxy for the position of the peak of the posterior (dotted lines). For DESI1.5, we find a positive neutrino mass at $\sum m_\nu^* \approx 0.03~\rm eV$, contrary to the previous DESI1 (FM) constraints, where the peak (according to our Gaussian approximation) lies on $\sum m_\nu^* \approx -0.09~\rm eV$. This is consistent with the parabolic fit to the profile likelihood performed in \cite{Y3.cpe-s2.Elbers.2025} that also finds a peak in the negative mass region. This suggests a shift of the whole marginalized posterior, which explains why the upper bound for DESI1 is more restrictive than for DESI1.5.

The right panel of the same figure shows the analogous posteriors when combining with CMB (\texttt{Camspec}) data and the CMB-only posterior for reference. The bottom panel of the same figure shows the 2-D contours in the $\Omega_m$, $\sum m_\nu$ space. We have added the BAO+CMB contours for reference to show that the constraints from FS+CMB and BAO+CMB do not differ much, due to the fact that the constraining power in the combination comes predominantly from the degeneracy breaking between the CMB and DESI contours. Nonetheless, we obtain 
\begin{equation}
    \sum m_\nu < 0.0624 ~\mathrm{eV} \quad (95\%\text{ CI}), \quad \text{DESI1.5 + CMB [\texttt{CamSpec}]},
\end{equation}
close to the analogous constraint from BAO+CMB \cite{DESI.DR2.BAO.cosmo}.

\subsubsection{\nuwowacdm{} constraints}
\begin{figure}
    \centering
    \includegraphics[width=0.49\linewidth]{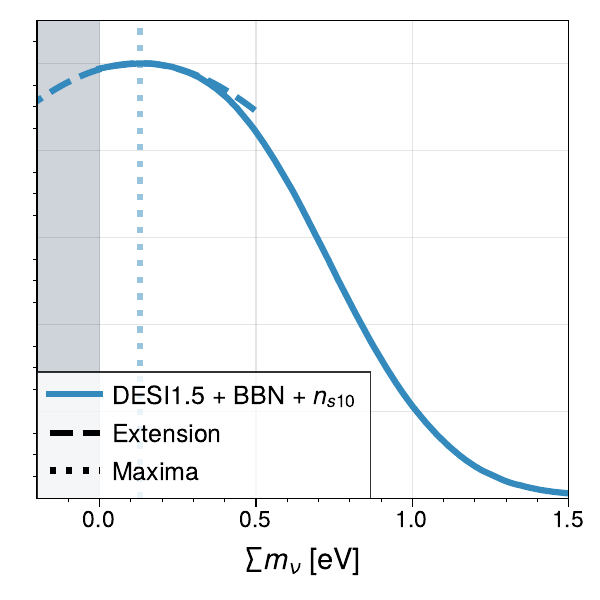}
    \includegraphics[width=0.49\linewidth]{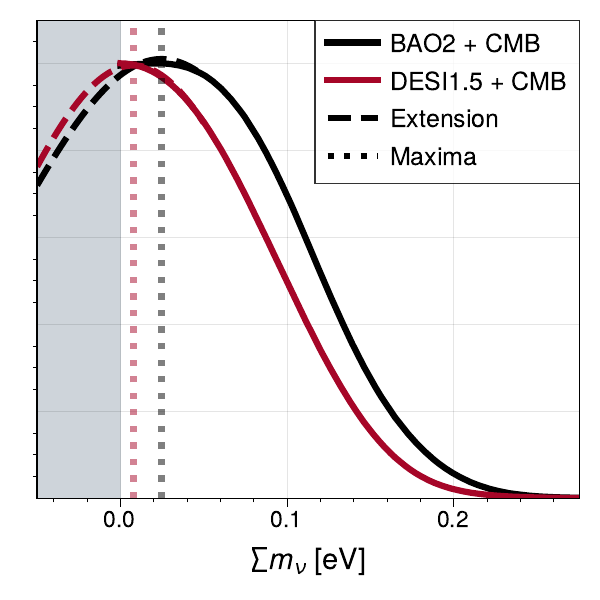} 
    \caption{Neutrino mass constraints within \wowacdm. Left: Marginalized 1-dimensional posterior for the sum of neutrino masses in a conservative DESI1.5+BBN+wide $n_s$ setup. Right: Marginalized 1-dimensional posterior on the sum of neutrino masses for BAO2+CMB and this work's DESI1.5+CMB. We extend the posteriors using a simple Gaussian fit (dashed lines) in order to find the location of their peak (dotted lines).}
    \label{fig:nuwowacdm}
\end{figure}
As in plain \wowacdm, our approach allows us to quote constraints using minimal additions to DESI data, for such a configuration, we obtain 
\begin{equation}
    \sum m_\nu < 1.004 ~\mathrm{eV} \quad (95\%\text{ CI}), \quad \text{DESI1.5 + BBN + $n_{s10}$}.
\end{equation}
Terrestrial particle experiments, such as KATRIN put model-independent constraints on the flavor-weighted mass of the electron neutrino $m_e^2 = \sum \abs{U_{e i}}^2m_i^2$, yielding an upper bound of $m_e < 0.31~\rm eV$\footnote{We cite the Feldman-Cousins constraint. We avoid drawing direct figure-to-figure comparisons due to the different interpretation of frequentist confidence levels and Bayesian confidence intervals.} at 90\% CL \cite{katrin}. In a naive, unrealistic scenario where we assume the lightest neutrino to be massless and the mixing matrix coefficients to be unity, this constraint can be roughly translated to a sum of neutrino masses of $\sum m_\nu \lesssim 0.93~\rm eV$ within the same frequentist framework. Our conservative DESI1.5 constraints are already comparable ($\sum m_\nu < 0.86~\rm eV$ at 90\% CI) even within a very flexible cosmological model. The left panel of \cref{fig:nuwowacdm} shows the marginalized posterior on the sum of neutrino masses in this scenario. We have also extended the posterior with a Gaussian fit to find the peak at $\sum m_\nu^* \approx 0.12~\rm eV$, which matches the maximum \textit{a posteriori} (MAP).

Combining with CMB, improves constraints by an order of magnitude 
\begin{equation}
    \sum m_\nu < 0.1444 ~\mathrm{eV} \quad (95\%\text{ CI}), \quad \text{DESI1.5 + CMB\texttt{[Camspec]}}.
\end{equation}
We obtain smaller constraints relative to BAO2+CMB, likely due to the posterior ``hiding'' back in the negative mass region. \Cref{fig:nuwowacdm} shows the marginalized posteriors on the sum of neutrino masses in a model with DE. In this case, we also extend the posteriors with a Gaussian fit and obtain that while both posteriors peak in the positive mass range, our DESI1.5 constraints peak very close to zero. Using our fit, we recover the value of $\sum m_\nu^* \approx 0.025~\rm eV$ for BAO + CMB obtained in \cite{Y3.cpe-s2.Elbers.2025} and obtain a value of $\sum m_\nu^* \approx 0.0078~\rm eV$ for our DESI1.5 constraints. The fact that the most likely neutrino mass shifts back towards zero supports our observation that our smaller upper bound is driven by an overall shift in the posterior rather than a large reduction in the uncertainty. %\DF{This is a very simple fit so no errors are estimated but not sure if it is worth the effort to get some.}

Finally, we observe that the dark energy constraints remain unchanged when freeing the neutrino mass, which is consistent with earlier observations using BAO data.

\subsubsection{\onulcdm{} constraints}

One last addition to neutrino constraints is relaxing flatness. In a spatially curved ($o-$) extension to \lcdm{} the curvature ``density'' $\Omega_k$ is allowed to vary. This is expected to change the distance-redshift relation, particularly increasing the distance to the last scattering surface and has been proposed \cite{Chen2025} as a possible solution to the BAO-CMB tension within \lcdm{} \cite{act2} and the neutrino mass tension.
\begin{figure}
    \centering
    \includegraphics[width=0.49\linewidth]{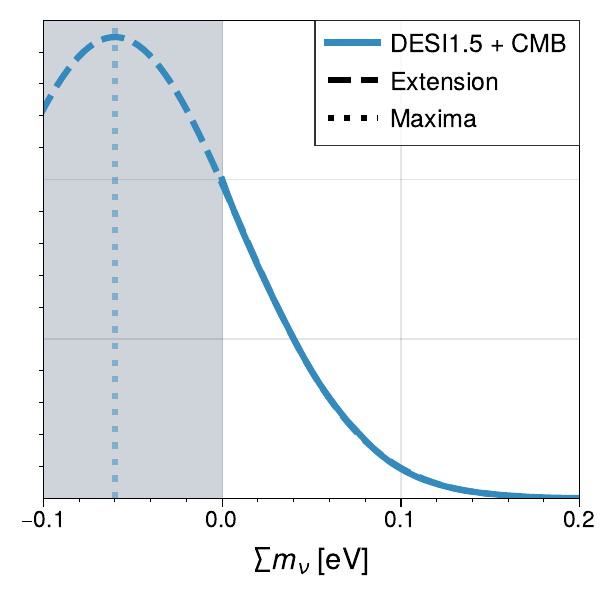}
    \includegraphics[width=0.49\linewidth]{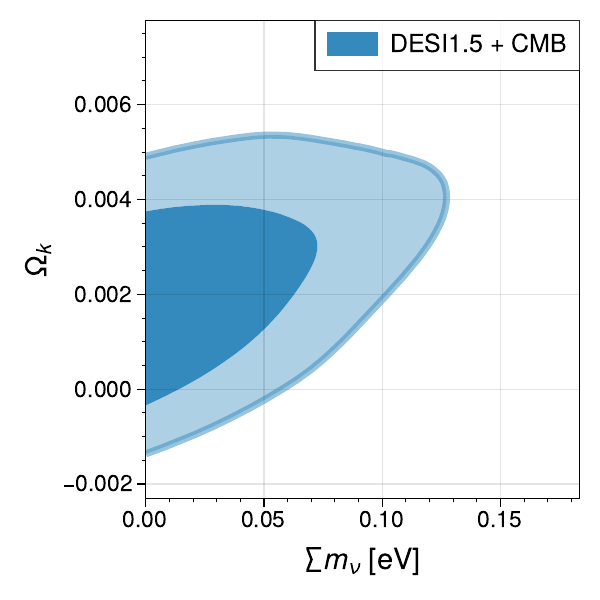}
    \caption{Marginalized posteriors in a non-flat cosmology with neutrinos. Here, we also show the Gaussian extension to the 1D neutrino mass posterior, showing its maximum in the negative region. The right panel shows the $\Omega_k-\sum  m_\nu$ plane, highlighting DESI1.5-only constraints.}
    \label{fig:nuomklcdm}
\end{figure}
\Cref{fig:nuomklcdm} shows the neutrino mass and curvature constraints from DESI1.5 combined with CMB. This combination should relax the neutrino mass constraints, given that the curvature freedom can account for some of the effect that a very small neutrino mass has on the transfer function.

For our baseline data combination, we obtain
\begin{equation}
    \sum m_\nu < 0.785 ~\mathrm{eV} \quad (95\%\text{ CI}), \quad \text{DESI1.5 + BBN + $n_{s10}$},
\end{equation}
which represents relaxed measurements relative to the flat case as predicted by \cite{Chen2025} but more constraining limits relative to the \wowacdm{} neutrino cosmology. 

We compute the Deviance Information Criterion (DIC) to find which cosmological model is favored in the presence of massive neutrinos. We find that $\Delta \rm DIC = -0.09$ between \lcdm{} and \olcdm{} and $\Delta {\rm DIC} = 3.72$ between \lcdm{} and \wowacdm{}, which suggests no preference of the data for either more complex model over \lcdm{} \cite{DIC}.

Adding CMB information, we obtain 
\begin{equation}
    \sum m_\nu < 0.100 ~\mathrm{eV} \quad (95\%\text{ CI}), \quad \text{DESI1.5 + CMB\texttt{[Camspec]}},
\end{equation}
which is also larger than the analogous \lcdm{} constraint and smaller than the \wowacdm{} one. In the former case, the posterior peaks at an effective value of $\sum m_{\nu,\rm eff}^* = -0.021~\rm eV$ and the freedom added by the curvature is not enough to make it move towards positive values. Using the same Gaussian-extension approach we used before, we find $\sum m_{\nu,\rm eff}^* = -0.060~\rm eV$, showing that in fact, this extra freedom does not particularly resolve the issue of a negative effective mass. In terms of DIC relative to \lcdm{}, the spatially curved and DE models show a $\Delta {\rm DIC} = -4.53$ and $\Delta {\rm DIC} = -0.20$, respectively, showing again no significant preference for either.

In terms of curvature, the presence of neutrino masses slightly alleviates the $\sim2\sigma$ tension of the data with a flat universe seen in the neutrinoless case (see \cref{sec:olcdm_constraints}). 

\subsection{Constraints on spatial curvature}
\label{sec:olcdm_constraints}

\begin{figure}
    \centering
    \includegraphics[width=0.49\linewidth]{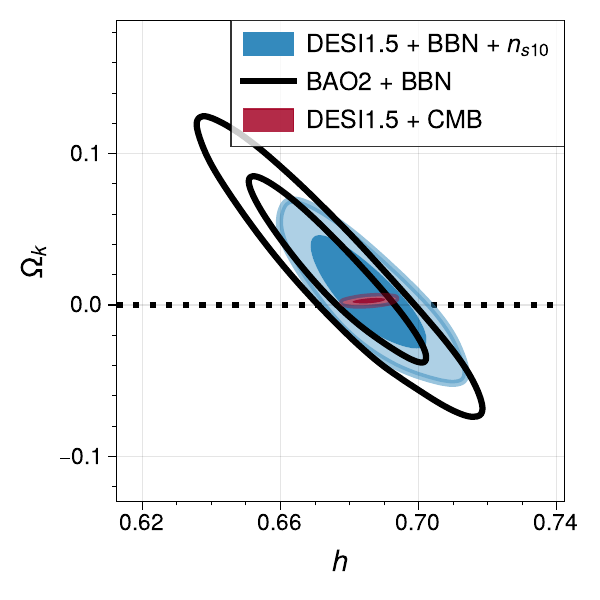}
    \includegraphics[width=0.49\linewidth]{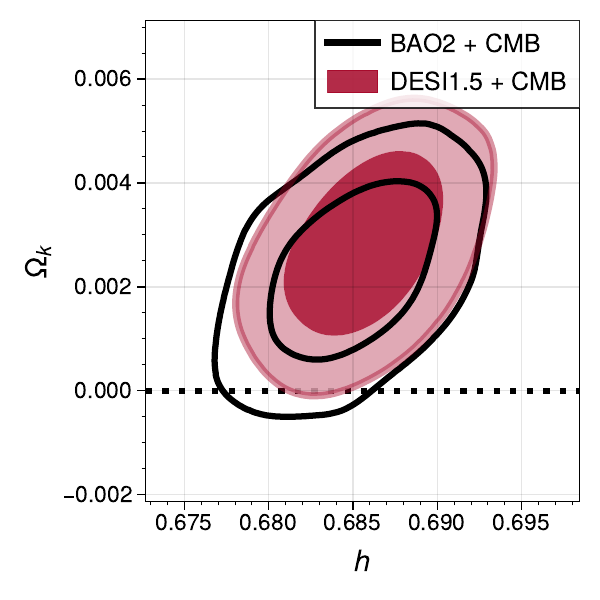}
    \caption{Marginalized posteriors in the $\Omega_k-h$ plane. We show our (DESI1.5) constraints combining with the conservative external data, and with CMB. The left panel shows BAO2 and DESI1.5 constraints without CMB. Right: Zoom-in plot to show DESI1.5+CMB constraints with the BAO2+CMB contours added for reference.}
    \label{fig:olcdm}
\end{figure}

Within DESI1.5, adding FS information to BAO2 tightens the constraint on the curvature parameter by a factor of two, while the geometry remains fully consistent with flatness, as can be seen on the left panel of \cref{fig:olcdm}. However, adding CMB information results in $\Omega_k > 0$ with a 95\% confidence or about 2.4$\sigma$ away from flatness. The right panel of \cref{fig:olcdm} presents the zoom-in on the CMB contours compared to the BAO-only constraints. These limits have been found to be slightly dependent on the CMB likelihood used: Ref.~\cite{Chudaykin2025b} has reported that using the HiLLiPoP+LoLLiPoP CMB likelihood decreases the significance of $\Omega_k > 0$ to below 2$\sigma$. Our measurements are summarized in \cref{tab:olcdm}.
\begin{table}
    \centering
    \resizebox{\columnwidth}{!}{\begin{tabular}{p{3cm}ccccc}
        \toprule
        Dataset & $\Omega_m$ & $h$ & $\Omega_k$ & $\sigma_8$ & $n_s$\\
\midrule 
\rowcolor[HTML]{EFEFEF}
BAO2 + BBN & $0.294 \pm 0.012$  & $0.676 \pm 0.017$  & $0.023 \pm 0.041$  & -- & --\\ 
\textbf{DESI1.5 + BBN + $n_{s10}$} & \boldmath$0.3042 \pm 0.0090$  & \boldmath$0.685 \pm 0.011$  & \boldmath$0.007 \pm 0.024$  & \boldmath$0.826 \pm 0.037$  & \boldmath$0.970 \pm 0.039$ \\ 
\rowcolor[HTML]{EFEFEF}
BAO2 + CMB & $0.3035 \pm 0.0038$  & $0.6850 \pm 0.0033$  & $0.0023 \pm 0.0011$  & $0.8141 \pm 0.0057$  & $0.9649 \pm 0.0040$ \\ 
\textbf{DESI1.5 + CMB} & \boldmath$0.3035 \pm 0.0036$  & \boldmath$0.6857 \pm 0.0030$  & \boldmath$0.0028 \pm 0.0011$  & \boldmath$0.8154_{-0.0055}^{+0.0051}$  & \boldmath$0.9639 \pm 0.0037$ \\ 
        \bottomrule
                        \end{tabular}}
                        \caption{{\bf Cosmological parameter constraints for the $o\Lambda$CDM model}, for different and data combinations: without CMB data (top rows); with CMB data (bottom rows), and where the BAO2 and DESI1.5 DESI dataset selections are displayed.}
                        \label{tab:olcdm}
                    \end{table}

\begin{figure}
    \centering
    \includegraphics[width=0.99\linewidth]{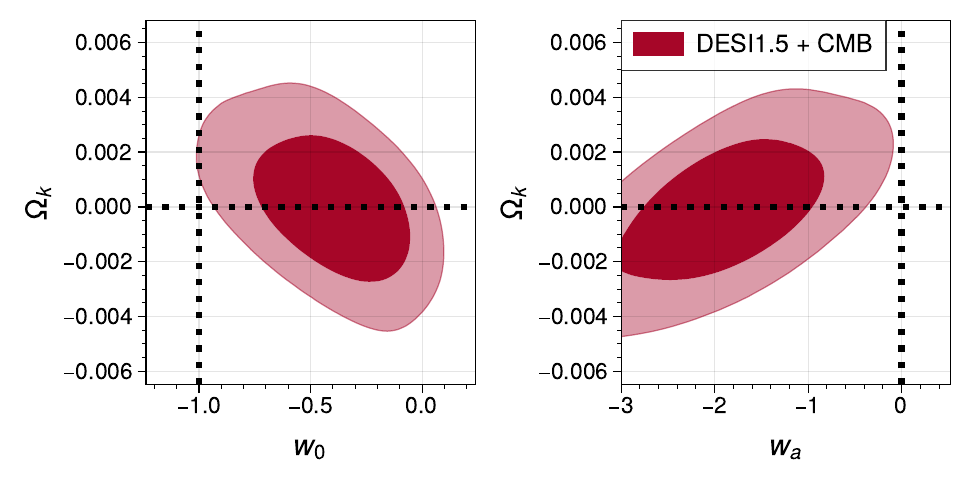}
    \caption{Marginalized posteriors in the $\Omega_k-w_a/w_0$ plane, for the \owowacdm{} model using DESI1.5 data in combination with CMB.}
    \label{fig:owowacdm}
\end{figure}

For completeness, we also explore the constraints of a spatially curved universe in combination with the $w_0$-$w_a$ parametrization for evolving dark energy. This is compactly shown in \cref{fig:owowacdm}, where we focus only on the degeneracy of $\Omega_k$ with these equation-of-state parameters. The 1D posteriors in this evolving dark energy \owowacdm{} model are,
\begin{equation}
    \begin{rcases}
    \Omega_m &= 0.354 \pm 0.022 \\
\Omega_k &= 0.0001_{-0.0017}^{+0.0015} \\
w_0 &= -0.41 \pm 0.23 \\
w_a &= -1.77 \pm 0.66 \\
\end{rcases}
    \quad
    \text{DESI1.5 + CMB},
    \end{equation}
perfectly consistent with a flat geometry, but with a small deviation away from the cosmological constant prediction, only slightly wider than the analogous \wowacdm{} constraints.

\subsection{Sound horizon-free constraints}
\label{sec:late_constraints}
We aim to perform an analysis assuming the $\Lambda$CDM model at only late epochs and being agnostic about the size of the sound horizon scale, $r_d$. Within the SF formalism, this is straightforward, as we only need to marginalize over an overall factor multiplying $\alpha_{\rm iso}$. In practice, this is done by introducing an extra nuisance parameter $r_{\rm fac}$ that modifies the effective sound horizon as $r_{d,\rm eff} = r_{\rm fac}r_d$. Refs.~\cite{Brieden2023,rdagnostic} have shown that, by introducing such a parameter and allowing it to vary, the BAO data no longer constrain $H_0$ while preserving information on $\Omega_m$. This factor affects all redshifts the same way, thus preserving the relative BAO sizes which encode the latter.

This approach allows us to perform an $H_0$ measurement independent of the sound-horizon calibration scale, which exploits the matter-radiation equality scale $k_{\rm eq}$ as a standard ruler to calibrate the distances within the model.
%BAO alone Using shape information,  the matter-radiation equality scale $k_{\rm eq} \propto \Omega_m h^2$ can be
%extracted from the clustering data. In combination with the BAO constraint on $\Omega_m$ this yields a measurement of $H_0$.
Within the SF formalism, this information is clearly linked to the $m$ parameter, which defines the slope at a pivot scale of $\sim0.03\,h{\rm Mpc}^{-1}$. In this case, we still need to include the BBN prior on $\omega_b$ to inform $m$ on the baryon suppression. 

Following the procedure stated above and matching the ones presented in~\cite{Brieden2023}, we find that the sound-horizon-free constraints for the DESI1.5+BBN+$n_{s10}$ within a late-time $\Lambda$CDM are given by,

\begin{equation}
    \begin{rcases}
    \Omega_m &= 0.302 \pm 0.011 \\
\sigma_8 &= 0.823 \pm 0.035 \\
h &= 0.693 \pm 0.026 \\
\end{rcases}
    \quad
    \text{DESI1.5 + BBN + $n_{s10}$.} 
    \end{equation}

The $h$ constraint (at the 3.7\% level) is as precise as those obtained by \cite{rdagnostic}, but the new BAO2 information shifts the mean by 0.8$\sigma$ to a lower value, bringing it in excellent agreement with the CMB $h$ determination. We also find a roughly $1\sigma$ deviation in the matter density relative to CMB.

\begin{figure}
    \centering
    \includegraphics[width=0.99\linewidth]{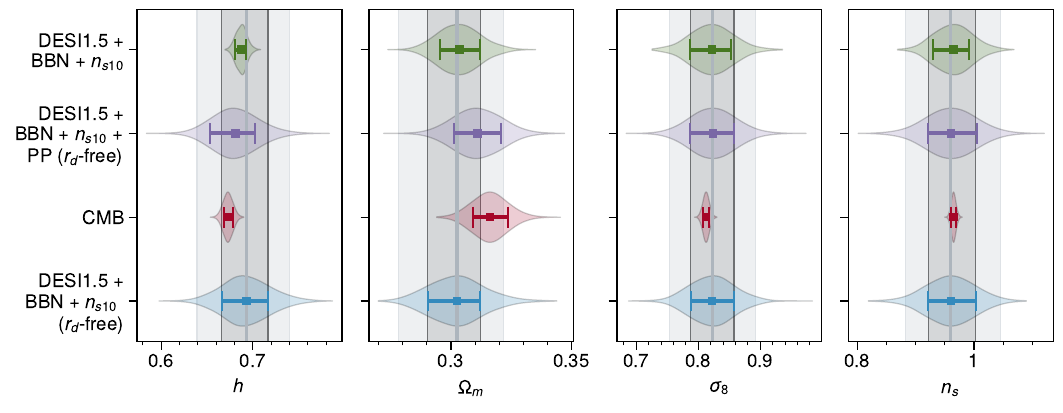}
    \caption{Constraints for sound-horizon-free ($r_d$-free) measurements with and without SNe (PP) compared to CMB and DESI1.5 measurements calibrated with the sound horizon (\cref{tab:cosmo_params_lcdm}) for comparison.}
    \label{fig:late_h}
\end{figure}

\Cref{fig:late_h}  displays this constraint in context, showing the $h$ CMB-only constraints, the DESI1.5 sound-horizon based constraints and sound-horizon free constraints also in combination with PP SNe sample. The SNe sample does not help to better constrain the parameters, but shifts the matter density by almost $0.5\sigma$ closer to the CMB value. %\DF{Not sure it makes much sense showing the SN, is the first time we show anything with SN and it seems it doesn't help much}

\begin{table}
    \centering
    \label{tab:cosmo_params}
    \scalebox{0.7}{\begin{tabular}{p{3cm}cccc}
        \toprule
        Dataset & $\Omega_m$ & $h$ & $\sigma_8$ & $n_s$\\
\midrule 
\rowcolor[HTML]{EFEFEF}
\textbf{DESI1.5 +
 BBN + $n_{s10}$
 ($r_d$-free)} & \boldmath$0.302 \pm 0.011$  & \boldmath$0.693 \pm 0.026$  & \boldmath$0.823 \pm 0.035$  & \boldmath$0.960 \pm 0.041$ \\ 
\textbf{DESI1.5 +
 BBN + $n_{s10}$
 + PP ($r_d$-free)} & \boldmath$0.3110 \pm 0.0097$  & \boldmath$0.681 \pm 0.025$  & \boldmath$0.824 \pm 0.036$  & \boldmath$0.961 \pm 0.042$ \\ 
        \bottomrule
\end{tabular}}
\caption{Sound-horizon-free cosmological parameter constraints from DESI1.5 within \lcdm{}. These constraints correspond to those shown in \cref{fig:late_h}.}
\end{table}

\section{Conclusions}
\label{sec:conclusions}

We have presented a consistent and optimal combination of the DESI DR1 Full-Shape and DR2 BAO measurements. By working in the compressed parameter space of the ShapeFit method, we have combined these datasets while accurately accounting for their correlations using a suite of EZmocks reprojected onto the DESI DR2 footprint. This approach, which we denote DESI1.5, achieves two main goals. First, it updates the DR1 Full-Shape constraints with the improved statistical power of DR2 BAO. Second, it mitigates the prior volume effects that have previously limited Bayesian interpretations of DESI-only data beyond the \lcdm{} without relying on more restrictive priors than those used in the official analysis.

We find that the ShapeFit compression substantially reduces projection effects in the marginalized posteriors. This allows us to report, for the first time within a Bayesian framework, DESI-only constraints on extensions such as dynamical dark energy and massive neutrinos using only minimal external priors from BBN and a wide prior on the spectral index $n_s$. The resulting constraints are robust and prior-resilient, closely aligning with frequentist profile likelihood results.

Our combined analysis tightens constraints on key parameters relative to DR1 alone. In the context of \lcdm{} we obtain $\Omega_m = 0.035 \pm 0.00085$, $h = 0.6876 \pm 0.0059$ and $\sigma_8 = 0.822 \pm 0.034$. For the \wowacdm{} dynamical dark energy model, we find $w_0 = -0.49\pm0.25$ and $w_a = -1.52\pm0.77$, representing a $\sim30\%$ improvement over DR1 (ShapeFit-based). The combination reduces the shift relative to \lcdm{} to 1.4$\sigma$ compared to the BAO-only analysis. When combined with CMB, the preference for an evolving dark energy component increases to about 3$\sigma$.

We also set competitive limits on the sum of neutrino masses. Using DESI1.5 with BBN and a wide $n_s$ priors, we find $\sum m_\nu < 0.54~\rm eV$ (95\% CI), and a posterior that peaks in the positive mass regime at $\sum m_\nu^* \approx 0.03~\rm eV$ in \lcdm{}, contrary to what was observed in the DESI DR1 analysis. In \wowacdm{} we obtain $\sum m_\nu < 1.00~\rm eV$ and $\sum m_\nu^* \approx 0.0078~\rm eV$. These are among the first neutrino mass constraints derived from a Bayesian analysis of DESI data without CMB. Adding CMB tightens these bounds to $\sum m_\nu < 0.062~\rm eV$ and $\sum m_\nu < 0.144~\rm eV$, respectively. 

Furthermore, the inclusion of Full-Shape information in addition to  DR2 BAO improves constraints on spatial curvature by a factor of two. While DESI1.5 alone remains consistent with a flat universe, combining it with CMB data yields a preference for an open universe at approximately 2.4$\sigma$; we note that this result may vary with alternative CMB likelihood choices.

Finally, in a sound-horizon-free analysis designed to depend on physics before $z\approx3000$ and only late-time observations, we obtain a constraint on the Hubble parameter of $h = 0.693 \pm 0.026$ without relying on recombination physics, demonstrating the growing power of large-scale structure to constrain cosmology independently of CMB.

The methodology developed here, i.e combining data releases at the compressed parameter level using mock-based covariances, provides a reliable alternative to Full-Modeling that mitigates parameter projection effects while still adopting wide uniform priors on nuisance parameters. The ShapeFit compression makes it possible to obtain robust Bayesian inference on beyond \lcdm{} models, allowing more definitive tests of dark energy, spatial curvature and neutrino properties with Large-Scale-Structure. 

\section*{Data availability}
The data to reproduce the figures in this paper is available in the DESI \texttt{zenodo} repository \url{https://doi.org/10.5281/zenodo.18629072}. (Public after acceptance)

\section*{Acknowledgements}

DFS and HGM acknowledge support through the Consolidación Investigadora (CNS2023-144605) of the Spanish Ministry of Science and Innovation. HGM also acknowledges the support of the Ramón y Cajal (RYC-2021-034104).
Funding for this work was partially provided by the Spanish MINECO under project
PID2022-141125NB-I00 MCIN/AEI, and the “Center of Excellence Maria de Maeztu 2020-
2023” award to the ICCUB (CEX2019-000918-M funded by MCIN/AEI/10.13039/501100011033).

This material is based upon work supported by the U.S. Department of Energy (DOE), Office of Science, Office of High-Energy Physics, under Contract No. DE–AC02–05CH11231, and by the National Energy Research Scientific Computing Center, a DOE Office of Science User Facility under the same contract. Additional support for DESI was provided by the U.S. National Science Foundation (NSF), Division of Astronomical Sciences under Contract No. AST-0950945 to the NSF’s National Optical-Infrared Astronomy Research Laboratory; the Science and Technology Facilities Council of the United Kingdom; the Gordon and Betty Moore Foundation; the Heising-Simons Foundation; the French Alternative Energies and Atomic Energy Commission (CEA); the National Council of Humanities, Science and Technology of Mexico (CONAHCYT); the Ministry of Science, Innovation and Universities of Spain (MICIU/AEI/10.13039/501100011033), and by the DESI Member Institutions: \url{https://www.desi.lbl.gov/collaborating-institutions}. Any opinions, findings, and conclusions or recommendations expressed in this material are those of the author(s) and do not necessarily reflect the views of the U. S. National Science Foundation, the U. S. Department of Energy, or any of the listed funding agencies.

The authors are honored to be permitted to conduct scientific research on I'oligam Du'ag (Kitt Peak), a mountain with particular significance to the Tohono O’odham Nation.

\appendix

\section{ShapeFit best-fit choice and projection effects}
\label{appendix:sf-choice}
\begin{figure}
    \centering
    \includegraphics[width=0.49\linewidth]{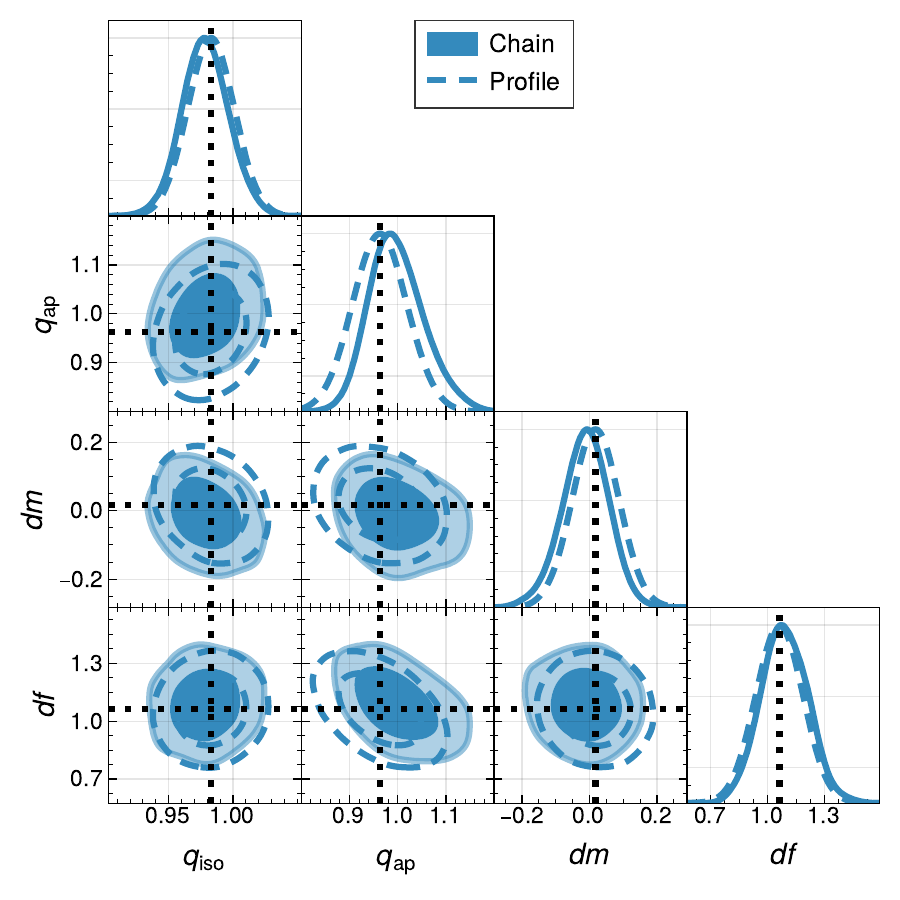}
    \includegraphics[width=0.49\linewidth]{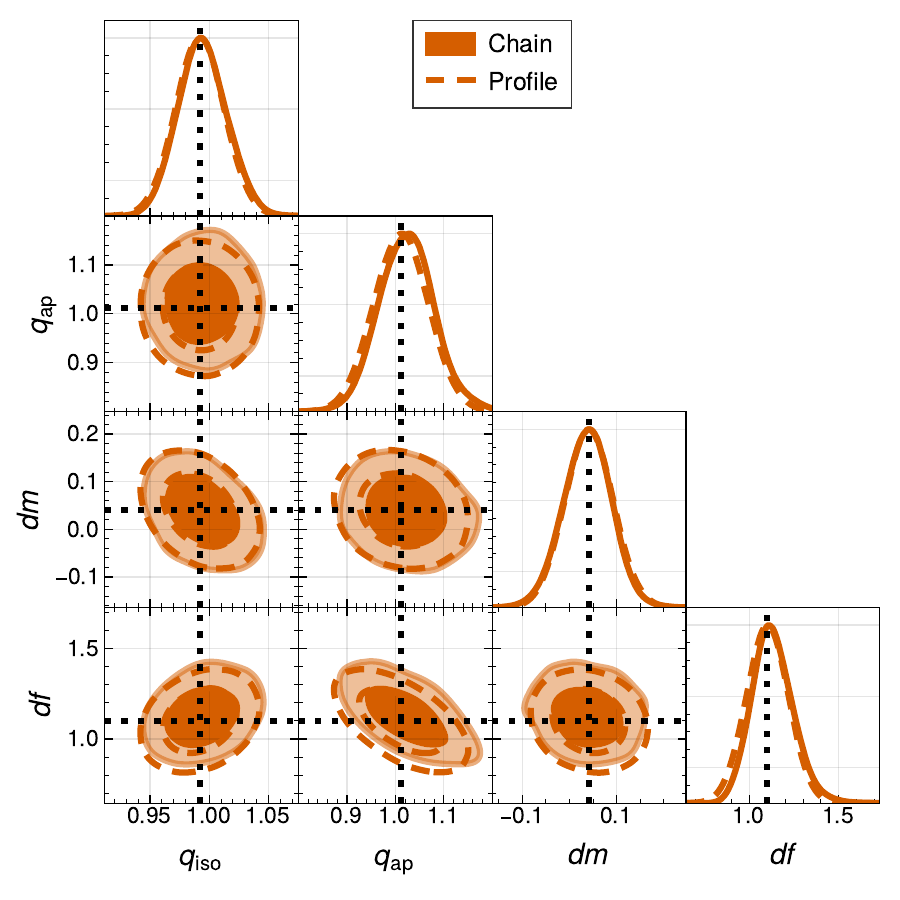}
    \caption{Marginalized posteriors of the ShapeFit compressed parameters for two galaxy samples: LRG1 on the left and QSO on the right. Filled contours show the MCMC Chains obtained from sampling the SF posterior, the empty dashed contours show the posterior profile as a multivariate Gaussian centered at the MAP values. Finally, the dotted marker lines show the approximate MAP value (i.e. taken from the MCMC chain, without profiling) used in this work. Both distributions are consistent with one another and the approximate and real MAP values coincide.}
    \label{fig:sf_posteriors}
\end{figure}
\begin{figure}
    \centering
    \includegraphics[width=0.99\linewidth]{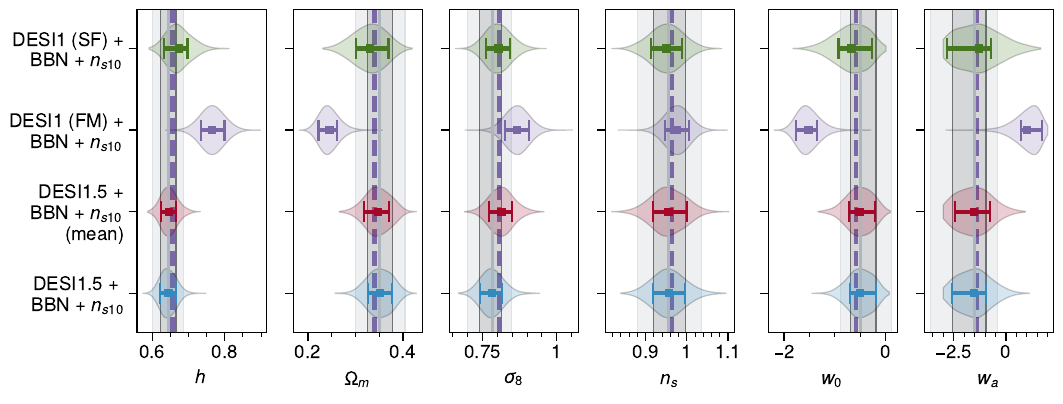}
    \caption{Marginalized posteriors for the DESI1.5 conservative fit within \wowacdm{} for two different choices of SF best-fit: the MAP (default used during this work) and the mean of the chain (mean). We add the DESI1 (FM) posteriors (purple) and their MAP (vertical dashed lines, as reported in \cite{Morawetz-frequentist}) and their analogous DESI1 (SF) posteriors. Vertical bands illustrate the 1 and 2$\sigma$ regions of the baseline presented in this paper.}
    \label{fig:sf-choice}
\end{figure}
In the SF interpretation step, we assume that the data is described by a multivariate Gaussian distribution, described by a mean $\mu^{\rm SF}$ and a covariance $\mathbf{C}^{\rm SF}$. If the SF posterior is exactly Gaussian, these mean and covariance are uniquely defined. However, if the posterior is non-Gaussian, we can choose to identify $\mu^{\rm SF}$ with the maximum or on the mean of the posterior.\footnote{Here we approximate the maximum of the posterior (MAP) as the point in the posterior chain with the maximum posterior. Rather than running an ensemble of minimizers on the posterior itself. This is a good approximation in the absence of projection effects.} 

Our baseline choice in the main text of the paper is to use the maximum. \Cref{fig:sf_posteriors} shows SF posteriors for two samples. The filled contours correspond to the actual MCMC posterior, whereas the dashed contours are a Gaussian approximation to the posterior obtained through a posterior profile and are thus centered at the real MAP values. The marker lines are the approximate MAP values used in this work. While for some samples, such as LRG1, shown on the left panel, there is a bias between MAP and the marginalized means, this difference is not significant, and the distributions remain consistent with one another to a great extent. For other tracers such as QSOs (on the right), the agreement is even better. This shows that the SF posteriors do not suffer from projection effects when marginalizing. Moreover, due to the agreement between the approximate and real MAP values, we do not expect to incur any biases due to our use of the approximate.

Finally, \cref{fig:sf-choice} shows the comparison of the marginalized posteriors of the cosmological interpretation of the DESI1.5 data under a \wowacdm{} model for the baseline choice (approximate MAP) and for the mean. Constraints from DESI1 (SF)  and (FM) are also shown for comparison. 
 Regardless of the choice for  $\mu^{\rm SF}$, both DESI1.5 posteriors show that projection effects (illustrated by the DESI1 (FM) posteriors and the corresponding MAP illustrated as vertical dashed lines) are largely mitigated. This is not unexpected as the SF-based posteriors are close to Gaussian as observed in \cref{fig:sf_posteriors}.

\section{Author Affiliations}
\label{sec:affiliations}

\noindent \hangindent=.5cm $^{a}${Institut de Ci\`encies del Cosmos (ICCUB), Universitat de Barcelona (UB), c. Mart\'i i Franqu\`es, 1, 08028 Barcelona, Spain.}

\noindent \hangindent=.5cm $^{b}${Departament de F\'{\i}sica Qu\`{a}ntica i Astrof\'{\i}sica, Universitat de Barcelona, Mart\'{\i} i Franqu\`{e}s 1, E08028 Barcelona, Spain}

\noindent \hangindent=.5cm $^{c}${Institut d'Estudis Espacials de Catalunya (IEEC), c/ Esteve Terradas 1, Edifici RDIT, Campus PMT-UPC, 08860 Castelldefels, Spain}

\noindent \hangindent=.5cm $^{d}${Instituci\'{o} Catalana de Recerca i Estudis Avan\c{c}ats, Passeig de Llu\'{\i}s Companys, 23, 08010 Barcelona, Spain}

\noindent \hangindent=.5cm $^{e}${University of Chinese Academy of Sciences, Nanjing 211135, People's Republic of China.}

\noindent \hangindent=.5cm $^{f}${Department of Astronomy, The Ohio State University, 4055 McPherson Laboratory, 140 W 18th Avenue, Columbus, OH 43210, USA}

\noindent \hangindent=.5cm $^{g}${The Ohio State University, Columbus, 43210 OH, USA}

\noindent \hangindent=.5cm $^{h}${Center for Cosmology and AstroParticle Physics, The Ohio State University, 191 West Woodruff Avenue, Columbus, OH 43210, USA}

\noindent \hangindent=.5cm $^{i}${Departamento de Astrof\'{\i}sica, Universidad de La Laguna (ULL), E-38206, La Laguna, Tenerife, Spain}

\noindent \hangindent=.5cm $^{j}${Instituto de Astrof\'{\i}sica de Canarias, C/ V\'{\i}a L\'{a}ctea, s/n, E-38205 La Laguna, Tenerife, Spain}

\noindent \hangindent=.5cm $^{k}${Lawrence Berkeley National Laboratory, 1 Cyclotron Road, Berkeley, CA 94720, USA}

\noindent \hangindent=.5cm $^{l}${Department of Physics, Boston University, 590 Commonwealth Avenue, Boston, MA 02215 USA}

\noindent \hangindent=.5cm $^{m}${Dipartimento di Fisica ``Aldo Pontremoli'', Universit\`a degli Studi di Milano, Via Celoria 16, I-20133 Milano, Italy}

\noindent \hangindent=.5cm $^{n}${INAF-Osservatorio Astronomico di Brera, Via Brera 28, 20122 Milano, Italy}

\noindent \hangindent=.5cm $^{o}${Centre for Astrophysics \& Supercomputing, Swinburne University of Technology, P.O. Box 218, Hawthorn, VIC 3122, Australia}

\noindent \hangindent=.5cm $^{p}${Department of Physics \& Astronomy, University College London, Gower Street, London, WC1E 6BT, UK}

\noindent \hangindent=.5cm $^{q}${Institute of Cosmology and Gravitation, University of Portsmouth, Dennis Sciama Building, Portsmouth, PO1 3FX, UK}

\noindent \hangindent=.5cm $^{r}${Institute of Space Sciences, ICE-CSIC, Campus UAB, Carrer de Can Magrans s/n, 08913 Bellaterra, Barcelona, Spain}

\noindent \hangindent=.5cm $^{s}${Institute for Computational Cosmology, Department of Physics, Durham University, South Road, Durham DH1 3LE, UK}

\noindent \hangindent=.5cm $^{t}${Instituto de F\'{\i}sica, Universidad Nacional Aut\'{o}noma de M\'{e}xico,  Circuito de la Investigaci\'{o}n Cient\'{\i}fica, Ciudad Universitaria, Cd. de M\'{e}xico  C.~P.~04510,  M\'{e}xico}

\noindent \hangindent=.5cm $^{u}${NSF NOIRLab, 950 N. Cherry Ave., Tucson, AZ 85719, USA}

\noindent \hangindent=.5cm $^{v}${University of California, Berkeley, 110 Sproul Hall \#5800 Berkeley, CA 94720, USA}

\noindent \hangindent=.5cm $^{w}${Institut de F\'{i}sica d’Altes Energies (IFAE), The Barcelona Institute of Science and Technology, Edifici Cn, Campus UAB, 08193, Bellaterra (Barcelona), Spain}

\noindent \hangindent=.5cm $^{x}${Departamento de F\'isica, Universidad de los Andes, Cra. 1 No. 18A-10, Edificio Ip, CP 111711, Bogot\'a, Colombia}

\noindent \hangindent=.5cm $^{y}${Observatorio Astron\'omico, Universidad de los Andes, Cra. 1 No. 18A-10, Edificio H, CP 111711 Bogot\'a, Colombia}

\noindent \hangindent=.5cm $^{z}${Fermi National Accelerator Laboratory, PO Box 500, Batavia, IL 60510, USA}

\noindent \hangindent=.5cm $^{aa}${Department of Astronomy, The University of Texas at Austin, 2515 Speedway Boulevard, Austin, TX 78712, USA}

\noindent \hangindent=.5cm $^{ab}${Institut d'Astrophysique de Paris. 98 bis boulevard Arago. 75014 Paris, France}

\noindent \hangindent=.5cm $^{ac}${IRFU, CEA, Universit\'{e} Paris-Saclay, F-91191 Gif-sur-Yvette, France}

\noindent \hangindent=.5cm $^{ad}${Department of Physics, The Ohio State University, 191 West Woodruff Avenue, Columbus, OH 43210, USA}

\noindent \hangindent=.5cm $^{ae}${Department of Physics, University of Michigan, 450 Church Street, Ann Arbor, MI 48109, USA}

\noindent \hangindent=.5cm $^{af}${University of Michigan, 500 S. State Street, Ann Arbor, MI 48109, USA}

\noindent \hangindent=.5cm $^{ag}${Department of Physics, The University of Texas at Dallas, 800 W. Campbell Rd., Richardson, TX 75080, USA}

\noindent \hangindent=.5cm $^{ah}${Department of Physics, Southern Methodist University, 3215 Daniel Avenue, Dallas, TX 75275, USA}

\noindent \hangindent=.5cm $^{ai}${Department of Physics and Astronomy, University of California, Irvine, 92697, USA}

\noindent \hangindent=.5cm $^{aj}${Institute of Physics, Laboratory of Astrophysics, \'{E}cole Polytechnique F\'{e}d\'{e}rale de Lausanne (EPFL), Observatoire de Sauverny, Chemin Pegasi 51, CH-1290 Versoix, Switzerland}

\noindent \hangindent=.5cm $^{ak}${Sorbonne Universit\'{e}, CNRS/IN2P3, Laboratoire de Physique Nucl\'{e}aire et de Hautes Energies (LPNHE), FR-75005 Paris, France}

\noindent \hangindent=.5cm $^{al}${Departament de F\'{i}sica, Serra H\'{u}nter, Universitat Aut\`{o}noma de Barcelona, 08193 Bellaterra (Barcelona), Spain}

\noindent \hangindent=.5cm $^{am}${Department of Physics and Astronomy, Siena University, 515 Loudon Road, Loudonville, NY 12211, USA}

\noindent \hangindent=.5cm $^{an}${Departamento de F\'{\i}sica, DCI-Campus Le\'{o}n, Universidad de Guanajuato, Loma del Bosque 103, Le\'{o}n, Guanajuato C.~P.~37150, M\'{e}xico}

\noindent \hangindent=.5cm $^{ao}${Instituto Avanzado de Cosmolog\'{\i}a A.~C., San Marcos 11 - Atenas 202. Magdalena Contreras. Ciudad de M\'{e}xico C.~P.~10720, M\'{e}xico}

\noindent \hangindent=.5cm $^{ap}${Department of Physics and Astronomy, University of Waterloo, 200 University Ave W, Waterloo, ON N2L 3G1, Canada}

\noindent \hangindent=.5cm $^{aq}${Perimeter Institute for Theoretical Physics, 31 Caroline St. North, Waterloo, ON N2L 2Y5, Canada}

\noindent \hangindent=.5cm $^{ar}${Waterloo Centre for Astrophysics, University of Waterloo, 200 University Ave W, Waterloo, ON N2L 3G1, Canada}

\noindent \hangindent=.5cm $^{as}${Instituto de Astrof\'{i}sica de Andaluc\'{i}a (CSIC), Glorieta de la Astronom\'{i}a, s/n, E-18008 Granada, Spain}

\noindent \hangindent=.5cm $^{at}${Departament de F\'isica, EEBE, Universitat Polit\`ecnica de Catalunya, c/Eduard Maristany 10, 08930 Barcelona, Spain}

\noindent \hangindent=.5cm $^{au}${Department of Physics and Astronomy, Sejong University, 209 Neungdong-ro, Gwangjin-gu, Seoul 05006, Republic of Korea}

\noindent \hangindent=.5cm $^{av}${CIEMAT, Avenida Complutense 40, E-28040 Madrid, Spain}

\noindent \hangindent=.5cm $^{aw}${Space Telescope Science Institute, 3700 San Martin Drive, Baltimore, MD 21218, USA}

\noindent \hangindent=.5cm $^{ax}${Department of Physics \& Astronomy, Ohio University, 139 University Terrace, Athens, OH 45701, USA}

\noindent \hangindent=.5cm $^{ay}${Department of Astronomy, Tsinghua University, 30 Shuangqing Road, Haidian District, Beijing, China, 100190}

\noindent \hangindent=.5cm $^{az}${National Astronomical Observatories, Chinese Academy of Sciences, A20 Datun Road, Chaoyang District, Beijing, 100101, P.~R.~China}

\bibliographystyle{JHEP} 
\bibliography{refs,DESI2024} 
\end{document}

%% file: authorlist.tex
\usepackage{orcidlink}
\affiliation{Affiliations are in Appendix \ref{sec:affiliations}}
\emailAdd{daniel.forerosanchez@icc.ub.edu}

\author[a]{{D.~Forero-Sánchez}\orcidlink{0000-0001-5957-332X},}
\author[b,c,a]{{H.~Gil-Mar\'in}\orcidlink{0000-0003-0265-6217},}
\author[d,a]{{L.~Verde}\orcidlink{0000-0003-2601-8770},}
\author[e]{{Z.~Ding}\orcidlink{0000-0002-3369-3718},}
\author[f,g,h]{{A.~J.~Ross}\orcidlink{0000-0002-7522-9083},}
\author[i,j]{{A.~Carnero Rosell}\orcidlink{0000-0003-3044-5150},}
\author[k]{{J.~Aguilar},}
\author[l]{{S.~Ahlen}\orcidlink{0000-0001-6098-7247},}
\author[k]{{S.~Bailey}\orcidlink{0000-0003-4162-6619},}
\author[m,n]{{D.~Bianchi}\orcidlink{0000-0001-9712-0006},}
\author[o]{{C.~Blake}\orcidlink{0000-0002-5423-5919},}
\author[k]{{A.~Brodzeller}\orcidlink{0000-0002-8934-0954},}
\author[p]{{D.~Brooks},}
\author[q]{{R.~Canning},}
\author[c,r]{{F.~J.~Castander}\orcidlink{0000-0001-7316-4573},}
\author[k]{{T.~Claybaugh},}
\author[s]{{S.~Cole}\orcidlink{0000-0002-5954-7903},}
\author[k]{{A.~Cuceu}\orcidlink{0000-0002-2169-0595},}
\author[t]{{A.~de la Macorra}\orcidlink{0000-0002-1769-1640},}
\author[u]{{Arjun~Dey}\orcidlink{0000-0002-4928-4003},}
\author[p]{{P.~Doel},}
\author[k,v]{{S.~Ferraro}\orcidlink{0000-0003-4992-7854},}
\author[d,w]{{A.~Font-Ribera}\orcidlink{0000-0002-3033-7312},}
\author[x,y]{{J.~E.~Forero-Romero}\orcidlink{0000-0002-2890-3725},}
\author[c,q,r]{{E.~Gaztañaga}\orcidlink{0000-0001-9632-0815},}
\author[z]{{G.~Gutierrez},}
\author[k]{{J.~Guy}\orcidlink{0000-0001-9822-6793},}
\author[aa]{{C.~Hahn}\orcidlink{0000-0003-1197-0902},}
\author[ab,ac]{{H.~K.~Herrera-Alcantar}\orcidlink{0000-0002-9136-9609},}
\author[h,ad,g]{{K.~Honscheid}\orcidlink{0000-0002-6550-2023},}
\author[ae,af]{{D.~Huterer}\orcidlink{0000-0001-6558-0112},}
\author[ag]{{M.~Ishak}\orcidlink{0000-0002-6024-466X},}
\author[u]{{R.~Joyce}\orcidlink{0000-0003-0201-5241},}
\author[u]{{S.~Juneau}\orcidlink{0000-0002-0000-2394},}
\author[ah]{{R.~Kehoe},}
\author[ai]{{D.~Kirkby}\orcidlink{0000-0002-8828-5463},}
\author[k]{{T.~Kisner}\orcidlink{0000-0003-3510-7134},}
\author[aj]{{J.~Kneib},}
\author[k]{{A.~Kremin}\orcidlink{0000-0001-6356-7424},}
\author[p]{{O.~Lahav}\orcidlink{0000-0002-1134-9035},}
\author[g]{{C.~Lamman}\orcidlink{0000-0002-6731-9329},}
\author[k]{{M.~Landriau}\orcidlink{0000-0003-1838-8528},}
\author[ak]{{L.~Le~Guillou}\orcidlink{0000-0001-7178-8868},}
\author[al,w]{{M.~Manera}\orcidlink{0000-0003-4962-8934},}
\author[u]{{A.~Meisner}\orcidlink{0000-0002-1125-7384},}
\author[d,w]{{R.~Miquel},}
\author[am]{{J.~Moustakas}\orcidlink{0000-0002-2733-4559},}
\author[an,ao]{{G.~Niz}\orcidlink{0000-0002-1544-8946},}
\author[ac,k]{{N.~Palanque-Delabrouille}\orcidlink{0000-0003-3188-784X},}
\author[ap,aq,ar]{{W.~J.~Percival}\orcidlink{0000-0002-0644-5727},}
\author[as]{{F.~Prada}\orcidlink{0000-0001-7145-8674},}
\author[at]{{I.~P\'erez-R\`afols}\orcidlink{0000-0001-6979-0125},}
\author[au]{{G.~Rossi},}
\author[av]{{E.~Sanchez}\orcidlink{0000-0002-9646-8198},}
\author[aw]{{E.~F.~Schlafly}\orcidlink{0000-0002-3569-7421},}
\author[k]{{D.~Schlegel},}
\author[ax]{{H.~Seo}\orcidlink{0000-0002-6588-3508},}
\author[k]{{J.~Silber}\orcidlink{0000-0002-3461-0320},}
\author[u]{{D.~Sprayberry},}
\author[af]{{G.~Tarl\'{e}}\orcidlink{0000-0003-1704-0781},}
\author[u]{{B.~A.~Weaver},}
\author[ay]{{C.~Zhao}\orcidlink{0000-0002-1991-7295},}
\author[k]{{R.~Zhou}\orcidlink{0000-0001-5381-4372},}
\author[az]{{H.~Zou}\orcidlink{0000-0002-6684-3997},}